\documentclass[final]{elsarticle}
\usepackage{framed,multirow}
\usepackage{amssymb}
\usepackage{amsmath}  % 提供 \text 命令
\usepackage{bm}
\usepackage{latexsym}
\usepackage[ruled,vlined]{algorithm2e} % 必须加载此包
\usepackage{enumitem}
\setlist{itemsep=2pt, topsep=3pt} % 减少条目间距
\usepackage{url}
\usepackage{xcolor}
\definecolor{newcolor}{rgb}{.8,.349,.1}
\usepackage{graphicx}          % 图形支持
\usepackage{wrapfig}
\usepackage[export]{adjustbox}
\usepackage{subcaption}        % 必须加载此包（替代 subfigure）
\usepackage{hyperref}
\usepackage{booktabs} % 必须加载此宏包以支持 \bottomrule 等命令
\usepackage{float}
\usepackage[switch,pagewise]{lineno} %Required by command \linenumbers below
\usepackage{stfloats}   % 双栏浮动体支持
\usepackage{tikz}
\usepackage{tikz-3dplot}

\journal{Computers \& Graphics}

\begin{document}

%\verso{Preprint Submitted for review}

\begin{frontmatter}

\title{A Remeshing Method via Adaptive Multiple Original-Facet-Clipping and Centroidal Voronoi Tessellation\tnoteref{tnote1}}%
\tnotetext[tnote1]{Only capitalize first
word and proper nouns in the title.}

\author[1]{Yue Fei}
\author[1]{Jingjing Liu}
\author[2]{Yuyou Yao}
\author[1]{Yusheng Peng}    
\author[1]{Liping ZHENG\corref{cor1}}

\cortext[cor1]{Corresponding author: 
	Tel.: +0-000-000-0000;  
	fax: +0-000-000-0000;}
\emailauthor{example@email.com}{Corresponding Author Name}

\address[1]{School of Computer Science and Information Engineering, Hefei University of Technology, Hefei 230601, China}

\address[2]{Anhui University of Science and Technology, Hefei 231131, China}

%\received{1 February 2017}
%\received{\today}
%%%% Do not use the below for submitted manuscripts
%\finalform{28 March 2017}
%\accepted{2 April 2017}
%\availableonline{15 May 2017}
%\communicated{S. Sarkar}

\begin{abstract}
%%%
CVT (Centroidal Voronoi Tessellation)-based remeshing optimizes mesh quality by leveraging the Voronoi-Delaunay framework to optimize vertex distribution and produce uniformly distributed vertices with regular triangles. Current CVT-based approaches can be classified into two categories: (1) exact methods (e.g., Geodesic CVT, Restricted Voronoi Diagrams) that ensure high quality but require significant computation; and (2) approximate methods that try to reduce computational complexity yet result in fair quality. To address this trade-off, we propose a CVT-based surface remeshing approach that achieves balanced optimization between quality and efficiency through multiple clipping times of 3D Centroidal Voronoi cells with curvature-adaptive original surface facets. The core idea of the method is that we adaptively adjust the number of clipping times according to local curvature, and use the angular relationship between the normal vectors of neighboring facets to represent the magnitude of local curvature. Experimental results demonstrate the effectiveness of our method.
%%%%
\end{abstract}


\end{frontmatter}

\section{Introduction}
\label{intro}
Surface remeshing improves computational efficiency and optimizes model accuracy by refining mesh distribution and reducing complexity. It is widely applied in animation, game development, physics simulations, and computational fluid dynamics, offering specific benefits such as enhancing real-time rendering or ensuring numerical stability based on the requirements of each field. In recent years, numerous methods have been proposed to address various objectives in surface remeshing, including local approaches \cite{Khan2018SurfaceRW, GUO201949}, anisotropic methods \cite{https://doi.org/10.1111/cgf.13877},  optimization-based techniques \cite{7927461}, and CVT-based methods \cite{WANG201551}. \citet{Khan2020_Survey} provides a detailed review of these advancements, offering insights into their classification, strengths, limitations, and future research opportunities.

Centroidal Voronoi Tessellation (CVT) \cite{Du1999_CVT} is a partitioning technique where Voronoi sites are located at the centroids of their respective cells. The dualization of CVT generates triangular meshes \cite{Bruno2010_CVT, Chen2018_RVC}. To utilize CVT for generating surface remeshing model, researchers have explored the precise computation of the intersections between CVT cells and the original model \cite{Liu2009_CVT}. Additionally, others have extended the distance definitions in the Voronoi diagram. For example, \citet{RONG2011475} applied the CVT generation algorithm to spherical and hyperbolic spaces to improve the remeshing of 3D models. Another study \cite{Liu2017_GVD} used geodesic distances to construct the geodesic Voronoi diagram. Although these methods can produce exactly remeshed surface model, their computational efficiency is low due to the complexity of intersection calculation and precise distance computation.

Due to the inefficiency of methods for exactly generating surface remeshing models, many researchers are exploring approximate techniques to efficiently simulate 3D model surface. For instance, the VoroCrust method \cite{Abdelkader2020_VoroCrust} introduces auxiliary points on the boundary to create Voronoi cells, simulating the surface boundary. The Restricted Tangent Face (RTF) method \cite{Yao2023_RTF} uses CVT to approximate 3D surface boundary with tangent planes. Additionally, the Power Diagram based Restricted Tangent Face (PowerRTF) method \cite{https://doi.org/10.1111/cgf.14897} leverages the capacity-constrained properties of the Power diagram to adaptively implement surface remeshing. While these algorithms are well-designed and easy to implement, they may struggle with highly complex models that feature intricate curvature variations and details, such as hair-like structures and sharp features, which may lead to topological errors like overlapping facets and disconnected regions.

In order to achieve a balance between enhancing output mesh quality and computational complexity in CVT-based surface remeshing, we propose a method that iteratively clips 3D Centroidal Voronoi cells using curvature-adaptive original surface facets. We employ a curvature-adaptive approach to perform multiple clips on individual Voronoi cells, enabling better approximation of the original model. This approach allows selecting effective clipping facets and determining the appropriate number of clips. Additionally, GPU acceleration achieves parallel computation across individual Voronoi cells. The main contributions of the proposed method are as follows: 

\begin{itemize}\setlength{\itemsep}{0pt}
	\item We propose a curvature-adaptive method that adjusts the number of clipping times per Voronoi cell proportional to local curvature magnitude. Through multiple clipping passes, we generate polygonal faces that achieve high-quality model remeshing.
	
	\item We propose a neighborhood-ring-based search strategy that locates neighboring facets with the highest Voronoi cell clipping potential by initiating from vertices of the original facet hosting the sample.
\end{itemize}

The remainder of this paper is organized as follows: A brief review of several CVT-based surface remeshing methods is provided in \autoref{sect:related_work}. Prior knowledge on CVT is explained in \autoref{sect:preliminary}. Our surface remeshing method is introduced in \autoref{sect:our_method}. Experimental results are presented in \autoref{sect:experiments}. Finally, conclusions are drawn in \autoref{sect:conclusion}.

\begin{figure*}[h]
	\centering
	% 子图1：使用 subfigure 环境
	\begin{subfigure}{0.23\textwidth}
		\includegraphics[width=\linewidth]{figs/bunny_input.pdf}
		\caption{Input} % 子图标题（留空则不显示编号）
		\label{subfig:input}
	\end{subfigure}
	\hfill
	% 子图2
	\begin{subfigure}{0.165\textwidth}
		\includegraphics[width=\linewidth]{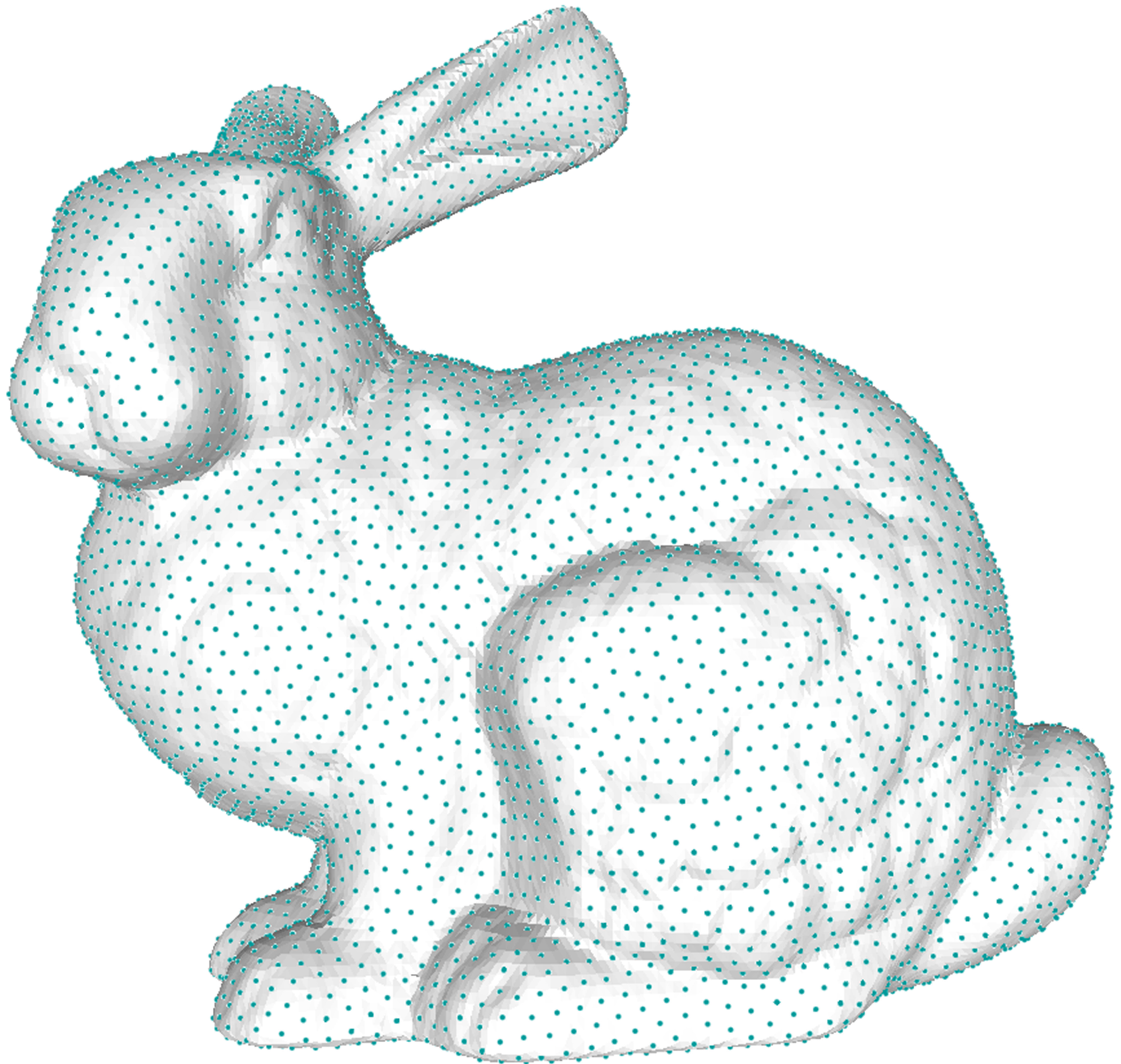}
		\caption{Initial sampling} % 子图标题
		\label{subfig:initial sample points}
	\end{subfigure}
	\hfill
	\begin{subfigure}{0.165\textwidth}
		\includegraphics[width=\linewidth]{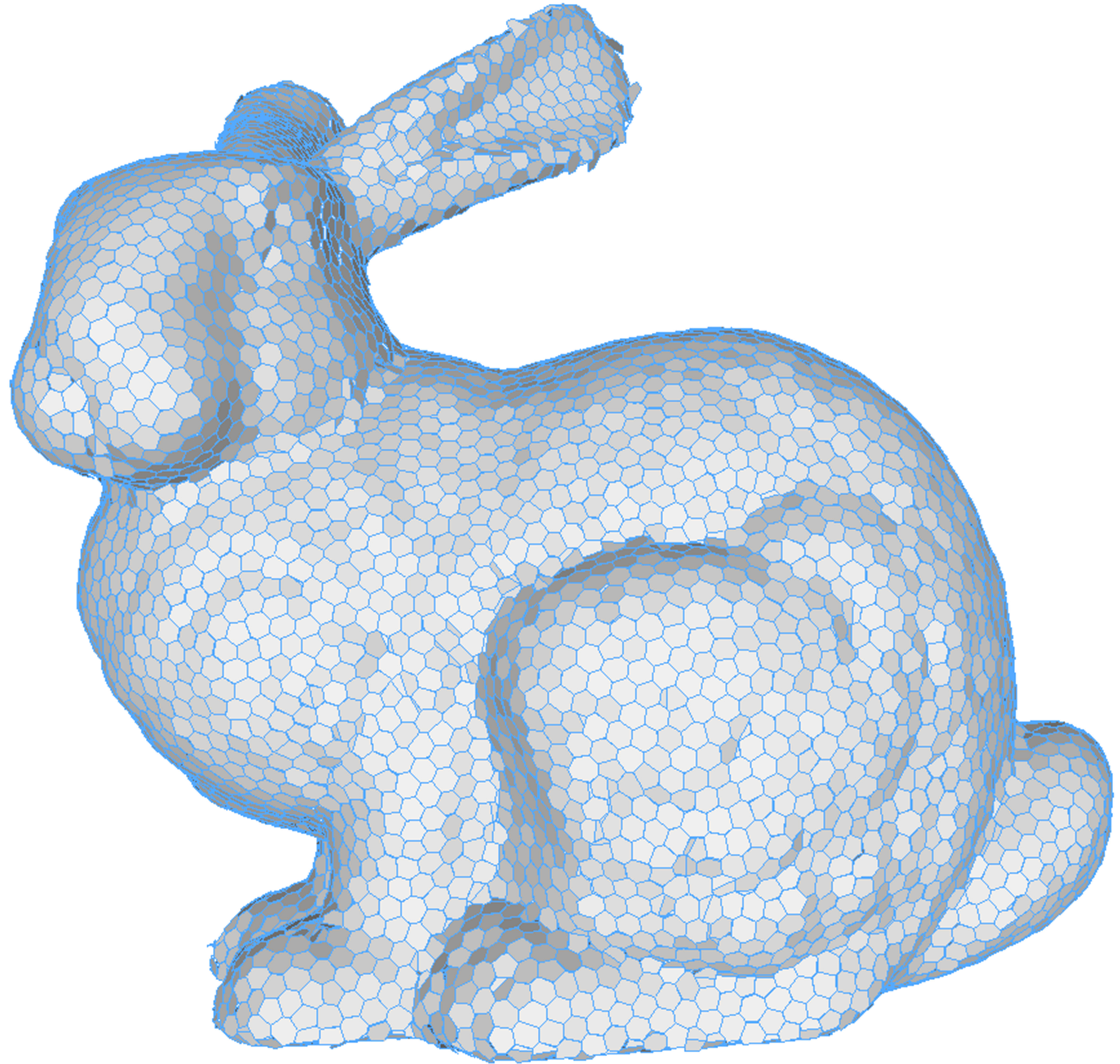}
		\caption{Initial clipped model} % 子图标题
		\label{subfig:initila clipped facets of model based initial points}
	\end{subfigure}
	\hfill
	\begin{subfigure}{0.165\textwidth}
		\includegraphics[width=\linewidth]{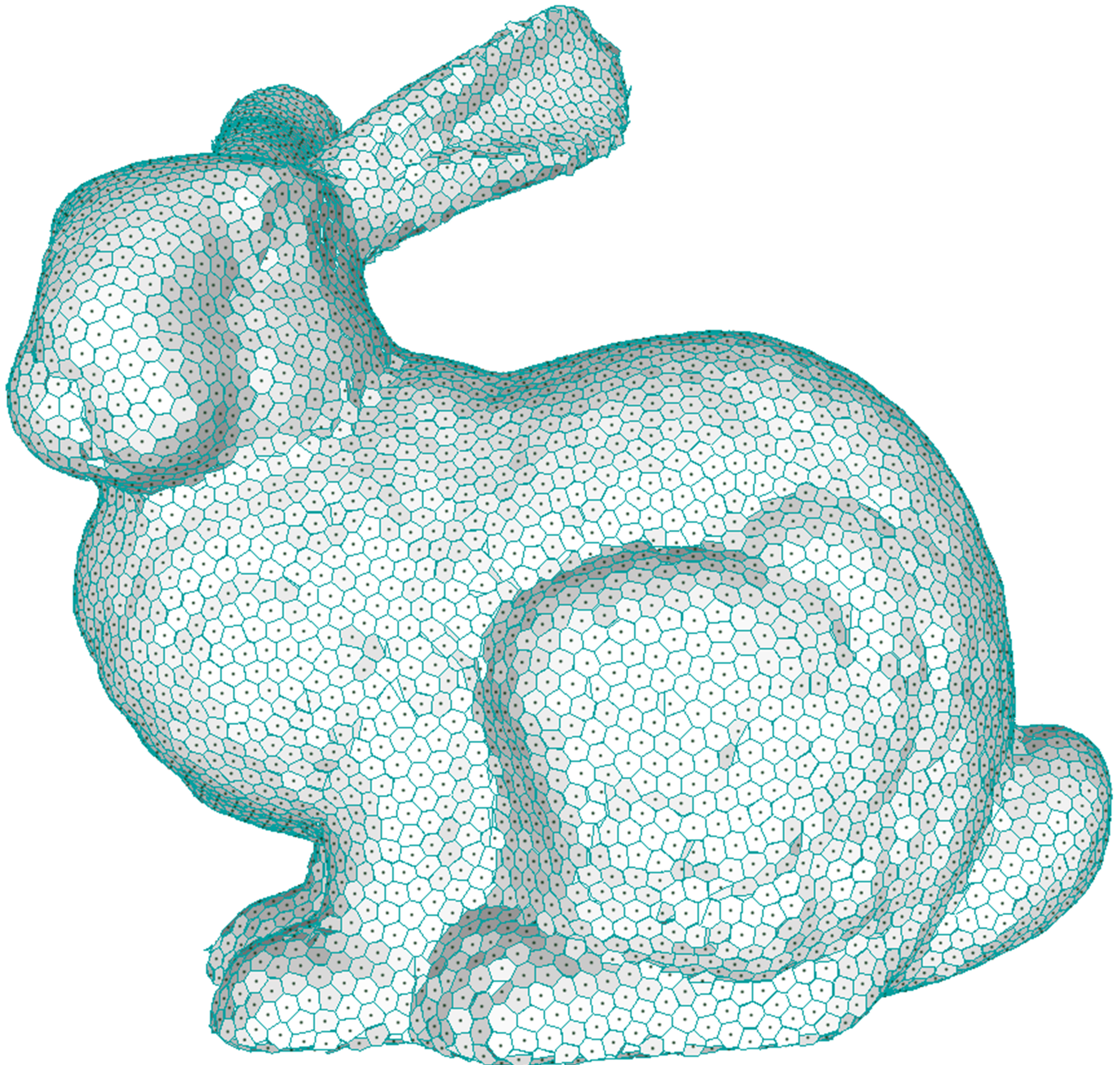}
		\caption{Optimized clipped model} % 子图标题
		\label{subfig:optimal clipped facets of model with multi times}
	\end{subfigure}
	\hfill
	\begin{subfigure}{0.25\textwidth}
		\includegraphics[width=\linewidth]{figs/bunny_output.pdf}
		\caption{Output} % 子图标题
		\label{subfig:output}
	\end{subfigure}
	\caption{Overall surface remeshing procedure with bunny by our method. 
		(a) Original model with 35.292k vertices and 70.580k facets, the average quality of its triangles is \textbf{0.715}; 
		(b) Initial sample points on original model; 
		(c) Clipped facets of model based on initial sample points; 
		(d) Optimal clipped facets of model after multiple times with optimized sample points; 
		(e) Output triangular mesh with 7k vertices and 13.996k facets, the average quality of its triangles is \textbf{0.917}. 
		The average quality of the output model has improved \textbf{28.252\%} compared to the original model. 
		Notably, the geometric details (e.g., leg-body junctions) exhibit higher-quality triangulation.
		}
	\label{fig:main}
	\vspace{-3pt}  % 更温和的间距调整
\end{figure*}

\section{Related work}
\label{sect:related_work}
Centroidal Voronoi Tessellation (CVT)-based methods, owing to their solid theoretical foundation and capability to enhance triangular mesh quality, have emerged as one of the pivotal methodologies in this field. For example, NASM \cite{NASM} proposes a GNN-driven high-dimensional normal metric CVT method, eliminating precomputed curvature fields and manual feature tagging, but it struggles with sparse-vertex CAD models due to training data constraints. CWF \cite{CWF} integrates CVT and normal anisotropy by incorporating decaying weights to balance accuracy, quality, and features, but faces challenges in anisotropic Restricted Voronoi Diagram (RVD) computation and efficiency. This section focuses on introducing relevant categories of CVT-based surface remeshing algorithms. For broader advancements in remeshing techniques, refer to the survey literature\cite{Khan2020_Survey}.

\subsection{Exact CVT-based remeshing}\label{sect:exact_cvt_based_remeshing}
The main challenge in exact CVT-based remeshing lies in the iterative computation of Voronoi diagrams on complex mesh surfaces. This process involves numerical optimization techniques to minimize CVT energy, thereby ensuring evenly distributed sample points \cite{Liu2009_CVT}. Exploring the relationship between meshes and their orthogonal duals enables the generation of high-quality triangular meshes through the duals of Voronoi diagrams formed by these sample points \cite{Mullen2011_HOT}. 

Existing methods for computing the exact Voronoi diagram on a surface can be broadly categorized into two main approaches. The first type involves computing the Restricted Voronoi Diagram (RVD) \cite{Yan2009_RVD}, which determines the intersection between Voronoi cell and the model surface. It accurately captures the boundaries of 3D model surfaces, resulting in high-quality remeshed surfaces. Subsequent research, such as that detailed by \cite{article}, further utilizes RVD to enhance high-dimensional surface remeshing techniques, significantly expanding the applications of RVD across various fields to improve mesh quality. For example, \citet{YAN2013843} improved the computational efficiency of RVD and successfully applied it to 3D closed models. Addressing challenges in non-contiguous regions, \citet{Yan2014_LRVD} introduced LRVD, while \citet{Yan2015_NonObtuse} added constraints to RVD for generating triangular meshes with minimal obtuse angles. Additionally, \citet{Du2018_FRVD} optimized degrees to further refine mesh quality, and \citet{Wang2020_RVDThinPlate} developed a robust algorithm tailored for closed thin-sheet models using RVD. Furthermore, \citet{Hou2022_SDFRVD} extended RVD to incorporate signed distance fields, showcasing its versatility in advancing meshing techniques. 

The second type involves extending the distance definition to construct more accurate Voronoi diagrams in surface, such as computing geodesic distances to form geodesic Voronoi diagrams. Methods such as GCVT \cite{Ye2019_GCVT} and GVD \cite{doi:10.1080/15481603.2023.2171703}\cite{Meng2023AnEA} have been developed for this purpose. Researchers have employed GVD to construct Delaunay triangulations (DT) for remeshing applications, as demonstrated in \cite{10.1145/2980179.2982424} and \cite{10.1145/2999532}. While exact CVT-based methods produce excellent results, they are associated with considerable computational intensity.

\subsection{Approximative CVT-based remeshing}
In order to alleviate the computational intensity in CVT-based remeshing, research has focused on approximate approaches to generating remeshed models. These approximate methods generally fall into two categories. The first type employs tangent plane methods. For instance, \citet{Zimmer2013_TPI} introduces additional degrees of freedom and deformed intersection formulas to better fit tangent planes to the model surface. \citet{Chen2018_RVC} proposed an approach involving clipped planes and Restricted Voronoi Cell (RVC) construction through site position resampling optimization for remeshing. Moreover, \citet{Chen2018_RVC} further extends CVT computation by integrating density functions to achieve density-adapted RVC (D-RVC). Notably, \citet{Yao2023_RTF} introduces the RTF algorithm, which employs plane cutting to approximate 3D model boundaries through tangent plane constraints. Furthermore, \citet{https://doi.org/10.1111/cgf.14897} extends Voronoi diagrams to Power diagrams and proposes the PowerRTF algorithm for adaptive curvature-aware remeshing. These methods offer advantages in terms of implementation simplicity and robustness, although their applicability is constrained to specific types of models.

The second approach involves using auxiliary points for approximative CVT-based remeshing, which typically incorporates clever design principles. For instance, the RPF method \cite{Xu2019_RPF} constructs Voronoi diagrams on input model boundaries via shadow points, enabling efficient computation and high-quality triangular mesh generation. The VoroCrust method \cite{Abdelkader2020_VoroCrust}, a recent advancement in approximate CVT-based remeshing, robustly constructs Voronoi cells along model boundaries by selecting circle centers and using their intersections as CVT sites. Overall, these auxiliary point-based methods achieve efficient performance with reduced algorithmic complexity, although their implementation requires significant geometric and mathematical expertise.

Our method also belongs to the category of approximate approaches, but it acts as a compromise between general approximate methods and exact methods, aiming to achieve high-quality remeshing while avoiding excessive computational time.

\section{Centroidal Voronoi Tessellation}
\label{sect:preliminary}
The Voronoi Tessellation \cite{Thiessen1911PRECIPITATIONAF}, originally introduced by Alfred H. Thiessen, serves as a method for partitioning regions to compute average rainfall. Its primary definition involves partitioning the given domain $\Omega$ surrounding a set of sample points $V(S)$, where each partition corresponds to the nearest sample point ${s}_i \in V(S)$ under the principle, with $||\cdot||$ denoting the Euclidean distance:
\begin{equation}
	V({s}_i) = \{{s} \in \Omega | \| {s} - {s}_i \| \leq \| {s} - {s}_j \|, \forall j \neq i \}
	\label{eq:vd}
\end{equation}

Adding centroid constraints to the Voronoi Tessellation results in the Centroidal Voronoi Tessellation (CVT)\cite{Liu2009_CVT}, which is defined as follows:

\begin{equation}
	{s}_i = {s}_{i}^* = \frac{\int_{V({s}_i)} {s}\rho({s})d{s}}{\int_{V({s}_i)} \rho({s})d{s}}
	\label{eq:cpd}
\end{equation}
where ${s}_i^*$ is the center-of-mass of the corresponding Voronoi cell, ${s}_i$ is each point, and ${s}_i$ is positioned at ${s}_i^*$. The $C^1$\mbox{-}smooth density function on $\Omega$ is defined as $\rho({s})$.

\begin{figure}[h]
	\centering
	% 子图1：使用 subfigure 环境
	\begin{subfigure}{0.355\linewidth}
		\includegraphics[width=\linewidth]{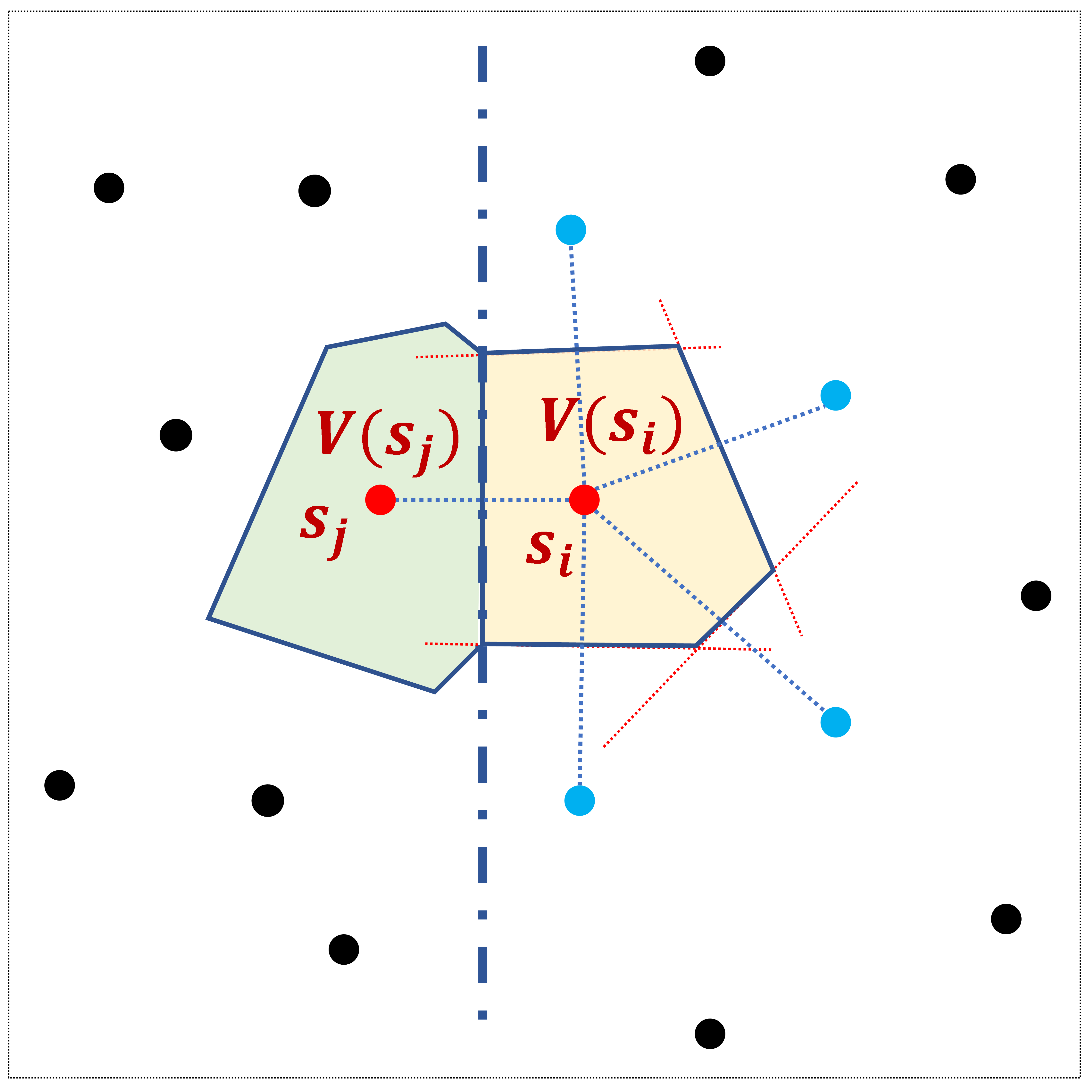}
		\caption{Voronoi Diagram} % 子图标题（留空则不显示编号）
		\label{fig:voronoi_diagram}
	\end{subfigure}
	\hfill
	% 子图2
	\begin{subfigure}{0.63\linewidth}
		\includegraphics[width=\linewidth]{figs/cvt.pdf}
		\caption{CVT} % 子图标题
		\label{subfig:CVT}
	\end{subfigure}
	\caption{Illustration of Voronoi Diagram and CVT.}
	\label{fig:cvt}
	\vspace{-3pt}  % 更温和的间距调整
\end{figure}

\section{Method}
\label{sect:our_method}
In this section, we present our method in detail, organized as follows: initial data processing is described in \autoref{subsect:method_inial}, followed by the adaptive clipping strategy in \autoref{subsect:method_clip}, centroid update computation in \autoref{subsect:method_centroid}, and final mesh extraction in \autoref{subsect:method_extract}.

\subsection{Overview}
\label{subsect:method_ini}
Let the input surface triangular mesh be defined as $$\mathbf{M} = (\mathbf{V}_{\mathbf{M}}, \mathbf{F}_{\mathbf{M}})$$
where:

$\mathbf{F}_{\mathbf{M}} = \{ f_t \}_{t=1}^{n_f}$ denotes the set of triangular faces, where each face $f_t$ is represented by an ordered triplet of vertex indices.

$\mathbf{V}_{\mathbf{M}} = \{ \bm{v}_{t,j} \mid t \in \{1, \dots, n_f\},\ j \in \{1,2,3\} \}$ denotes the ordered vertex coordinates, with $\bm{v}_{t,j}$ being the $j$-th vertex of the $t$-th face.
The workflow, which is further detailed in \autoref{algo:overview}, consists of four phases: (1) \textbf{initial sampling}, (2) \textbf{compute cell and clipping}, (3) \textbf{update sites}, and (4) \textbf{mesh extraction}. It is also illustrated in \autoref{fig:main}.

%我们的方法主要可分为四步：1在模型表面初始化采样，并计算每个原始面片的法向量；迭代2和3直至满足收敛条件2：根据站点位置计算CVT，对每一个cell进行多次的裁剪，3更新站点；4提取网格
  \begin{algorithm}[t]
	\small
	\caption{Surface remeshing using Original-Facet-Clipping and Centroidal Voronoi Tesssellation}
	\label{algo:overview}
	\KwIn{original surface $\mathbf{M}$, number of sampling points $n$,error termination $\epsilon$,  maximum number of iterations $N_{m}$ }
	\KwOut{triangular meshes $\mathbf{M^\prime}$}
  \tcp{Phase 1: Initialization}
	$\mathbf{S} = \{ \mathbf{s}_i \}_{i=1}^n$ are uniformly sampled from $\mathbf{M}$\\
	Calculate the normal $\bm{n}_t$ for each original mesh face\\
	Set the counter $n_{it}$ to 0 \\
	\tcp{Phases 2-3: Iterative Optimization}
	\While{$\delta > \epsilon$ and $n_{it} < N_{m}$}{
		\For{$\mathbf{s}_i \in \mathbf{S}$ in parallel}{  
			\tcp{Compute and clip the cell}
			Compute the Centroid Voronoi Cell $ V(\mathbf{s}_i) $ use Lloyd's method\\
			Construct the neighboring facet set $ F_{\text{near}}$ \\
			Determine the number of clipping counts\\
			Select original facets for clipping\\
			Clip the cell based frame work \cite{MeshlessVoronoiGPU} \\
			
			\tcp{Update position of site}
			Calculate the barycenter $b_i$ \\ 
			Calculate the projection point $b_i^p$ by projecting the barycenter $b_i$ on $\mathbf{M}$  \\
			${s}_i$ $\leftarrow$ $b_i^p$ \\
		}
		$n_{it} \leftarrow n_{it} + 1$
	}
	    \tcp{Phase 4: Mesh Extraction}
	Extract the triangular meshes
\end{algorithm}
\vspace{-2pt}  % 更温和的间距调整

\begin{wrapfigure}{r}{0.39\linewidth}
	\includegraphics[width=\linewidth]{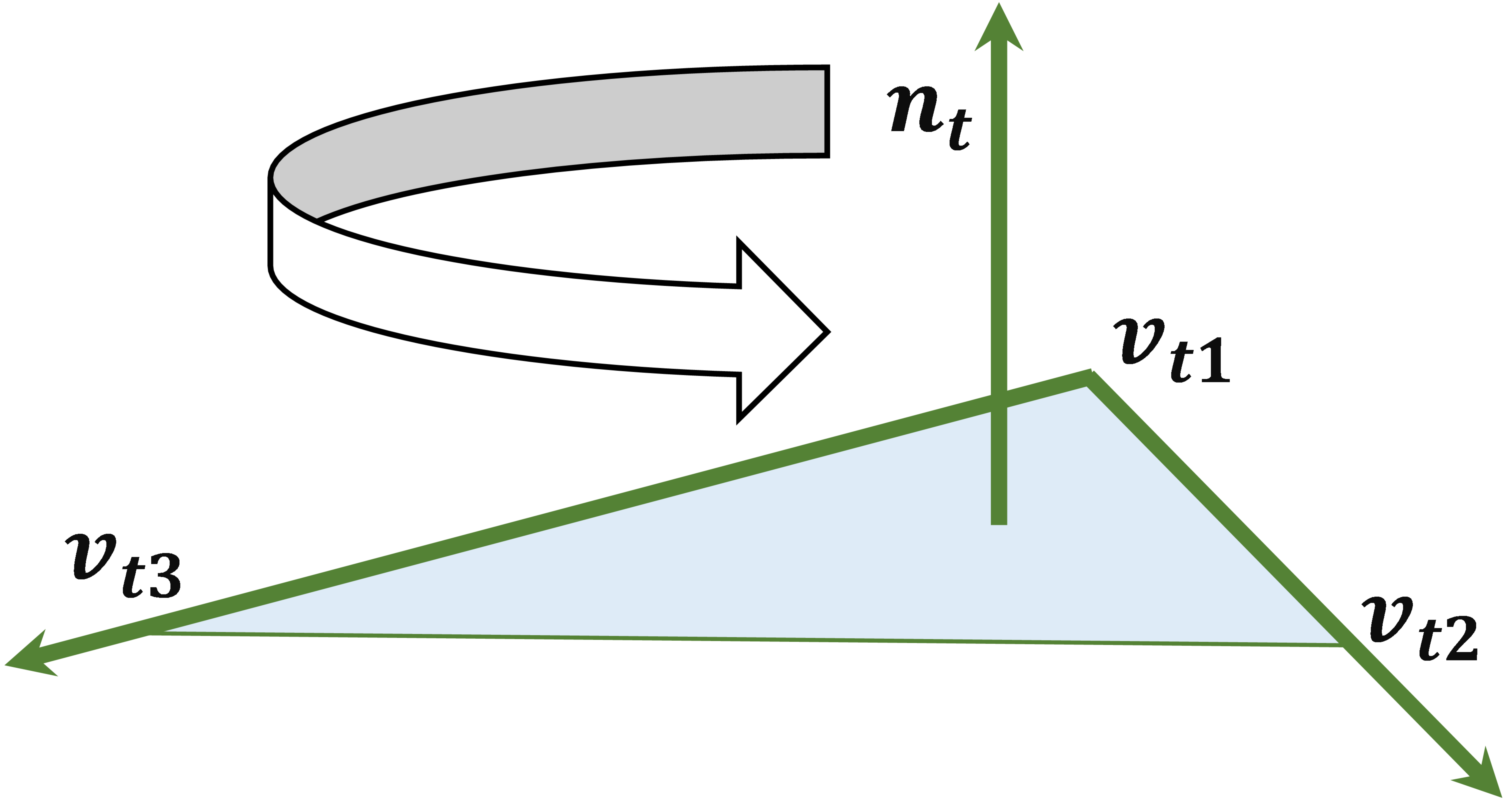}
	\vspace{-3pt}
\end{wrapfigure}
\subsection{Initialization}
\label{subsect:method_inial}
During this phase, we perform two primary operations: (1) a set of sample points $\mathbf{S} = \{ \mathbf{s}_i \}_{i=1}^n$ is generated by uniform surface sampling on $\mathbf{M}$ using the Geogram library \cite{Levy2015_Geogram}, and (2) for each original mesh face $ f_t $, the normal vector $\bm{n}_t$ is computed by applying the right-hand rule \autoref{eq:right_hand} to its three vertex coordinates.
\begin{equation}
	\bm{n}_t =  (\bm{v_{t3}} - \bm{v_{t1}}) \times (\bm{v_{t2}} - \bm{v_{t1}}) 
	\label{eq:right_hand}
\end{equation}

  \begin{figure*}[t!]
	\centering
	% 子图1：使用 subfigure 环境
	\begin{subfigure}{0.495\textwidth}
		\includegraphics[width=\linewidth]{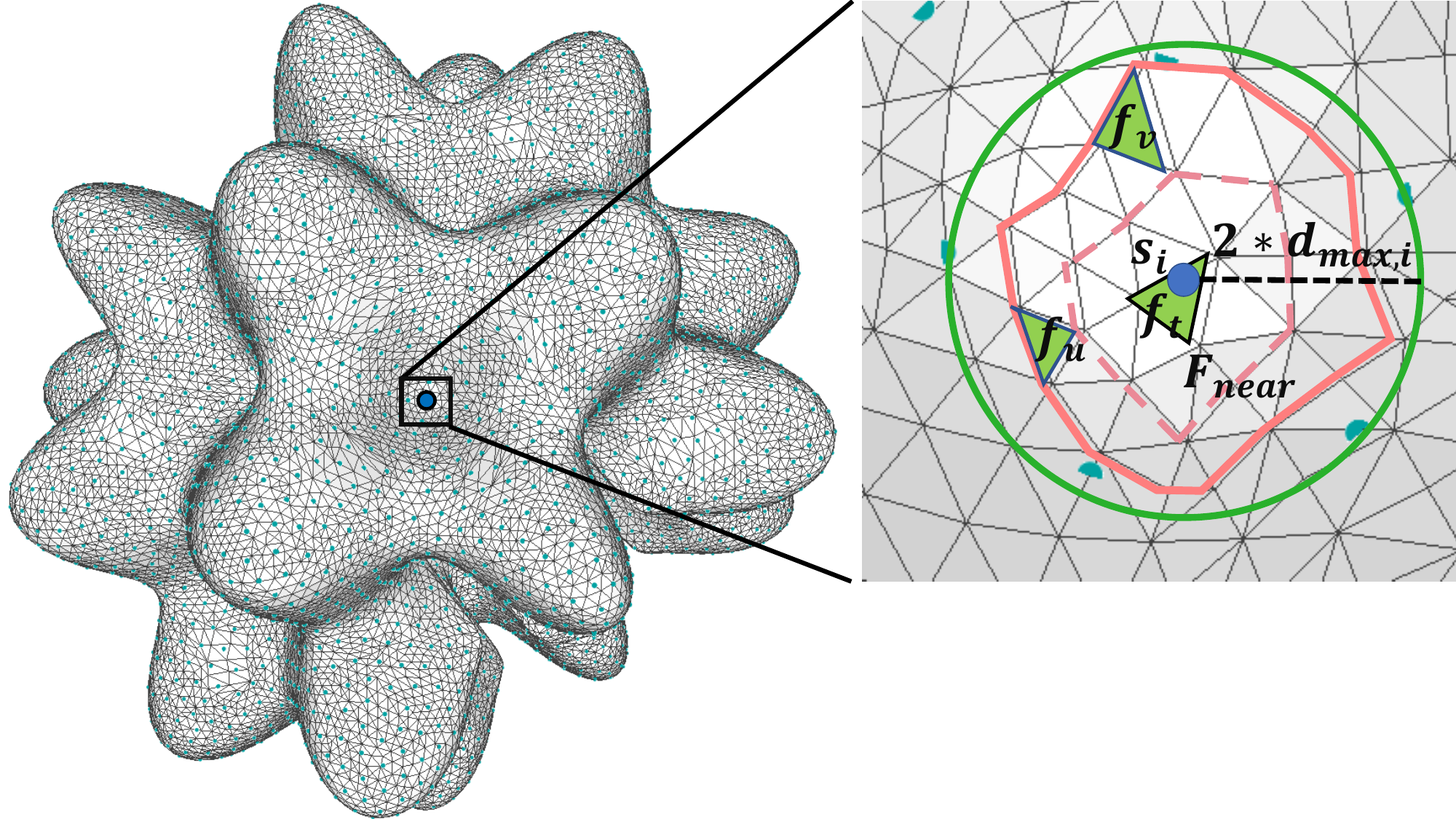}
		\caption{Construction of neighboring facet set} % 子图标题（留空则不显示编号）
		\label{subfig:Fnear}
	\end{subfigure}
	\hfill
	% 子图2
	\begin{subfigure}{0.495\textwidth}
		\includegraphics[width=\linewidth]{figs/clip_facet.pdf}
		\caption{The model with clipped cell (by original-facets)} % 子图标题
		\label{subfig:clip_facets}
	\end{subfigure}
	\caption{(a) The blue point is the current site $\mathbf{s}_i$, where: The red dashed line encloses the first-layer neighborhood-ring of the original triangular facet $\mathbf{f}_t$ containing $\mathbf{s}_i$;  The pink solid line encloses the second-layer neighborhood-ring of $\mathbf{f}_t$;  The green circle defines a range with radius $2 \times d_{\max,i}$ (where $d_{\max,i} = \max_{\mathbf{s}_k \in N(\mathbf{s}_i)} \|\mathbf{s}_k - \mathbf{s}_i\|$, and $N(\mathbf{s}_i)$ denotes the neighboring sites of $\mathbf{s}_i$). The set of original triangular facets lying within both neighborhood-rings and not exceeding the green circle is defined as the neighboring facet set $F_{\text{near}}$ of $\mathbf{s}_i$. By traversing facets in $F_{\text{near}}$ and calculating the cosine values of angles between each facet and $\mathbf{f}_t$, it is determined that the current cell requires three clippings. The clipping facets $\mathbf{f}_u$ and $\mathbf{f}_v$ are identified using \autoref{eq:score nd clip} and \autoref{eq:score th clip}. (b) After performing three clippings on the cell, three sub-facets are generated: $F_{i1}$ (light green), $F_{i2}$ (dark green), and $F_{i3}$ (deepest green).  }
	\label{fig:clip_strategy}
	\vspace{-3pt}  % 更温和的间距调整
\end{figure*}

\begin{figure}[b!]  % [!htbp] 建议使用更明确的浮动位置参数
	\centering
	\includegraphics[width=0.99\linewidth]{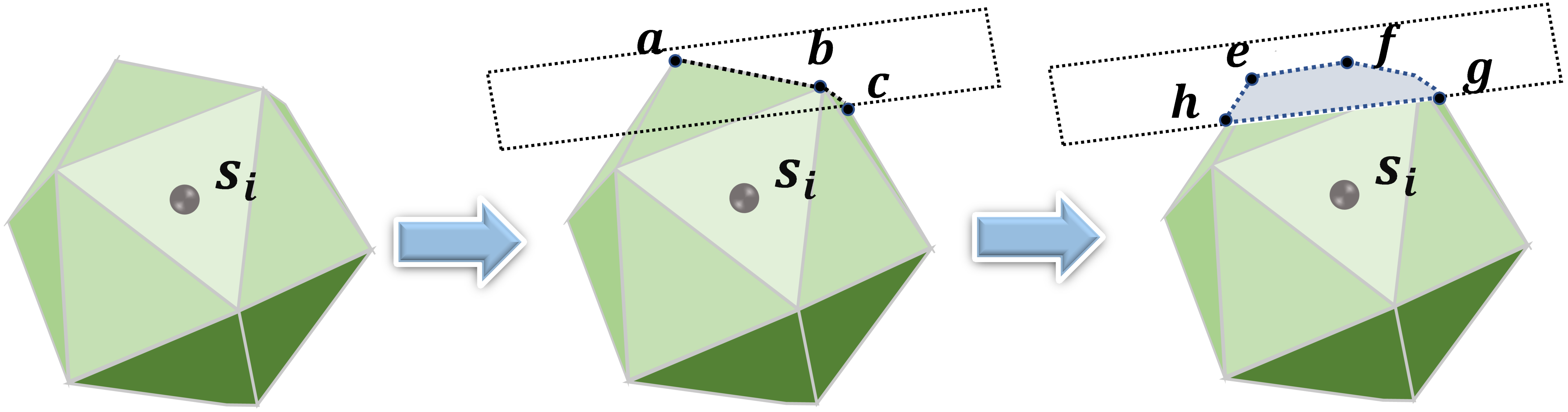} % 建议保留2%边距
	\caption{Half-space clipping process: First, we identify the removal side containing vertices $a$, $b$, $c$ and eliminate them with connected edges; Second, computing intersections $e$, $f$, $g$, $h$ between the clipping plane and cell edges and generating new facet by connecting these intersections.} % 分号分隔层级
	\label{fig:clip_a_cell}  % 正确位置
	\vspace{-6pt}  % 更温和的间距调整
\end{figure}

\subsection{Clipping Strategy}
\label{subsect:method_clip}
We adopt the parallelized clipping framework from \cite{MeshlessVoronoiGPU} for Voronoi cell computation (defined as $V(\mathbf{s}_i)$) and polyhedral clipping. The framework operates through two key mechanisms: (1) KNN-driven Voronoi computation, where the Voronoi cell of site $\mathbf{s}_i$ is iteratively clipped by bisecting planes $\Pi(\mathbf{s}_i, \mathbf{s}_k)$ generated by its $K$ nearest neighbors $\{\mathbf{s}_k\}_{k=1}^K$; (2) half-space partitioning, which defines the vertex set of the convex cell, identifies and removes vertices on the non-retained side, computes intersections between the half-plane and crossing edges, connects these points cyclically, and updates the vertex set. As also can be see in \autoref{fig:clip_a_cell}.

Common methods for computing Centroidal Voronoi Tessellation (CVT) include Lloyd's method and Newton's method. Although Newton's method offers faster convergence, it requires geometric continuity between Voronoi cells, which is disrupted by discontinuous gaps introduced through clipping operations in our method. We therefore adopt Lloyd's method, which imposes no continuity constraints and demonstrates stronger robustness.

\textbf{\textit{Constructing neighboring facet set.}} We set the clipping times for each site to range from 1 to 3 times, with three curvature levels defined as follows: Level 1 (minor curvature), Level 2 (intermediate curvature), and Level 3 (remarkable curvature). For each Centroid Voronoi cell $ V(\mathbf{s}_i) $, the number of clipping times is adaptively determined by the local curvature around its site $ \mathbf{s}_i $. Considering computational complexity and time constraints, we avoid explicit curvature calculations and instead approximate the relative curvature by measuring the maximum included angle between adjacent triangular facets of the input mesh. This angle is computed based on face normals generated during the initialization phase.

To achieve this, we first construct a neighboring facet set $ F_{\text{near}} $ for the original facet $ f_t $ where each site $ \mathbf{s}_i $ is located. This set is designed to cover all original facets potentially enclosed by the cell $ V(\mathbf{s}_i) $ , enabling both curvature estimation and facet selection for clipping operations through their half-planes. To reduce algorithmic complexity, we adopt a simplified neighbor set construction method: all facets within the twice-ring neighborhood of $ f_t $ are included to prevent undersized sets. However, since some twice-ring facets may be too far from $ f_t $ due to elongated one-ring facets, we constrain the maximum distance between the centroid of any facet in the set and the centroid of $ f_t $ to be no more than twice the distance to the farthest neighboring site $ \mathbf{s}_k $ used in the cell's computation. This effectively limits the search to original facets approximately enclosed by the cell. The detailed process can be referred to in \autoref{fig:clip_strategy}(\subref{subfig:Fnear}).

\textbf{\textit{Determining the number of clipping counts.}}
The angles between facets in $F_{\text{near}}$ and the original triangular facet $f_t$ can be assessed based on their face normals. If the absolute cosine values of these angles exceed a threshold $\bm{\alpha}$ (in our method the value is 0.8), Level 1 curvature (requiring one clipping) is indicated. Conversely, if any facet in $F_{\text{near}}$ has an absolute cosine value of the angle with $f_t$ below $\bm{\alpha}$, Level 2 curvature (necessitating at least two clippings) is identified.

\begin{figure}[t!]  % [!htbp] 建议使用更明确的浮动位置参数
	\centering
	\includegraphics[width=0.99\linewidth]{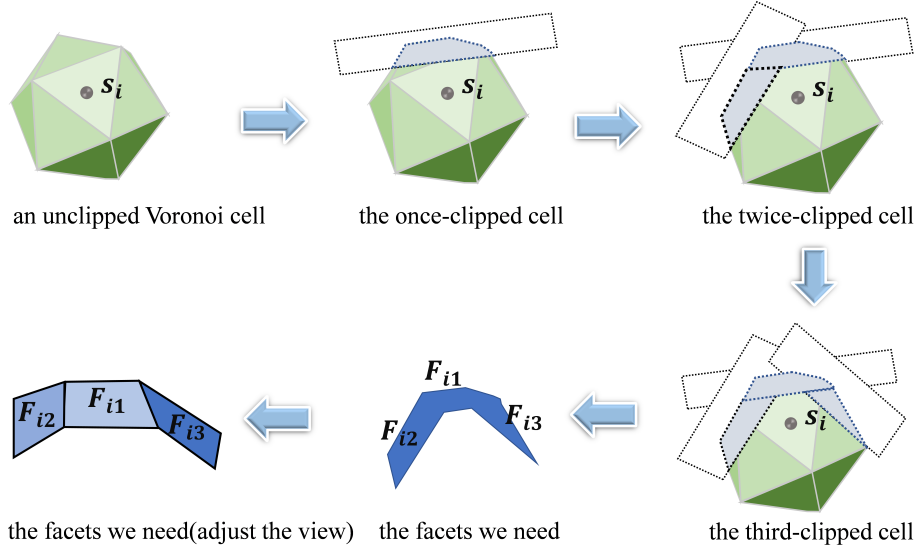} % 建议保留2%边距
	\caption{The Voronoi cell undergoes three sequential clipping operations along the arrow direction, starting from the unclipped state at the top-left and terminating at the bottom-right. The algorithm retains only the facets $F_{i1}$, $F_{i2}$, $F_{i3}$ generated through these clips. The middle facet in the bottom row represents the final retained result, while the leftmost element provides an adjusted viewpoint for intuitive facet inspection.} % 分号分隔层级
	\label{fig:clip_cell}  % 正确位置
	\vspace{-8pt}  % 更温和的间距调整
\end{figure}

If the Centroid Voronoi cell $ V(\mathbf{s}_i) $  requires at least two clippings and the second clipping facet is defined as $f_u$, further evaluation determines the need for a third clipping. For each triangular facet in $F_{\text{near}}$, angles relative to both $f_t$ and the second clipping facet $f_u$ are computed. If a facet in $F_{\text{near}}$ has absolute cosine values relative to both $f_t$ and $f_u$ below a threshold $\bm{\beta}$ (in our method the value is 0.7), significant curvature (Level 3) is confirmed, requiring a third clipping. The detailed process of a cell clipped three times can be referred to in \autoref{fig:clip_cell}, while the surface model with facets after three clippings is shown in \autoref{fig:clip_strategy}(\subref{subfig:clip_facets}).

\begin{figure}[b!]
	\centering
	% 子图1：使用 subfigure 环境
	\begin{subfigure}{0.47\linewidth}
		\includegraphics[width=\linewidth]{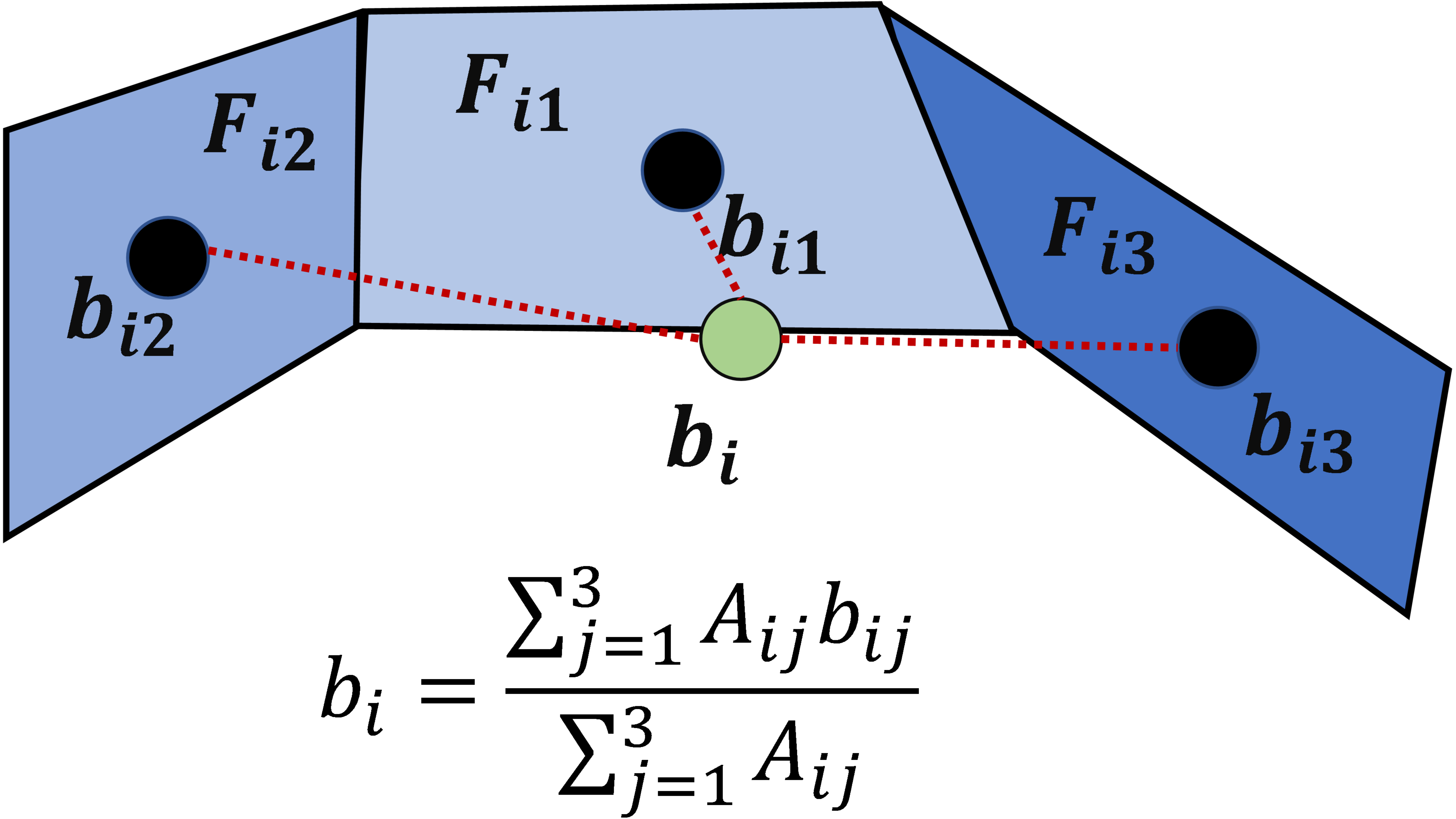}
		\caption{} % 子图标题（留空则不显示编号）
		\label{subfig:calculating_centroid_a}
	\end{subfigure}
	\hfill
	% 子图2
	\begin{subfigure}{0.5\linewidth}
		\includegraphics[width=\linewidth]{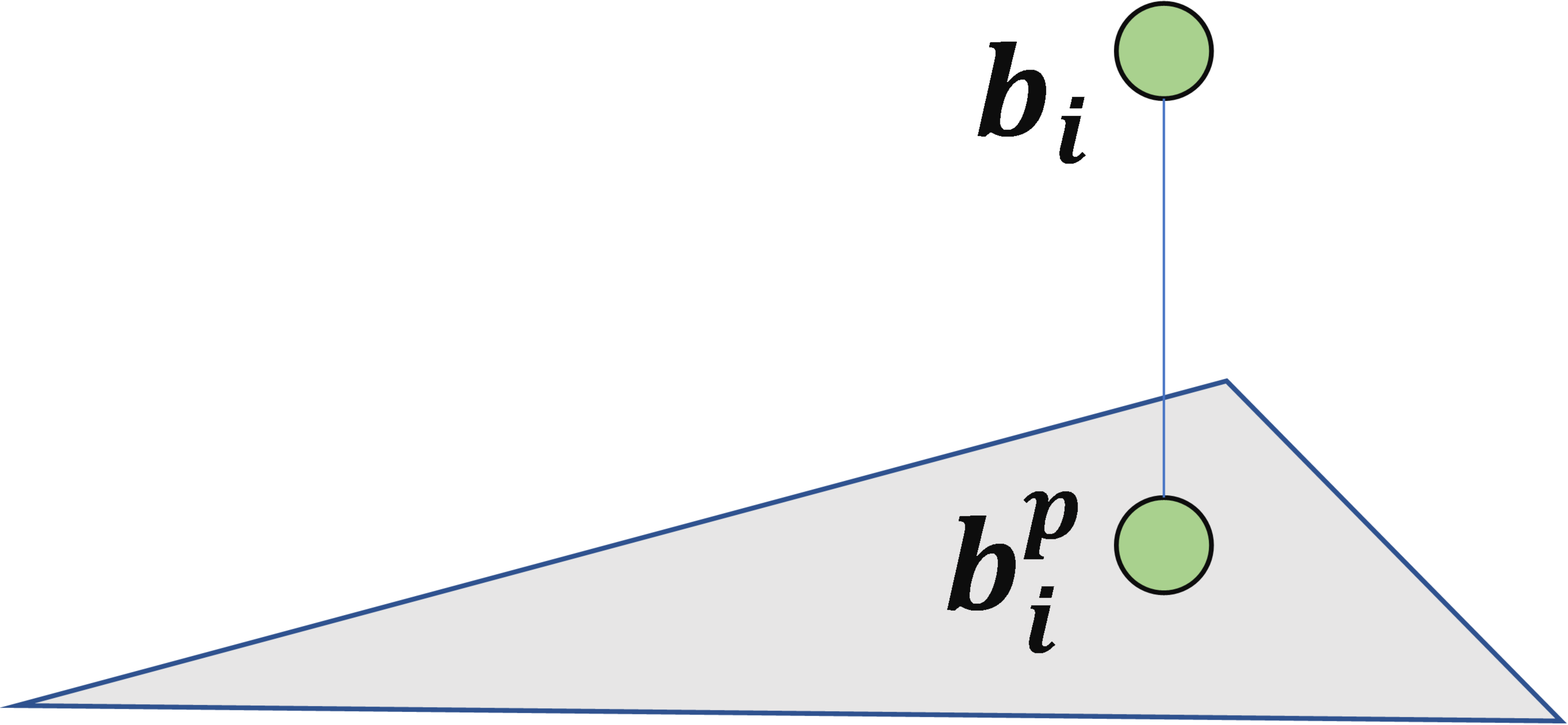}
		\caption{} % 子图标题
		\label{subfig:calculating_centroid_b}
	\end{subfigure}
	\caption{Illustration of calculating the centroid. (a) Compute the weighted centroid of multiple patches using \autoref{eq:cal_cent}, where m is set to 3 since three patches are involved;	 (b) If the calculated centroid lies outside the model surface, project it onto the original model surface following the strategy described in the \cite{Yao2023_RTF}.}
	\label{fig:calculating_centroid}
	\vspace{-8pt}  % 更温和的间距调整
\end{figure}

\textbf{\textit{Selecting original facets for clipping.}}
Based on the previously mentioned content, when a Centroid Voronoi cell $ V(\mathbf{s}_i) $ requires at least two clipping operations, all eligible triangular facets in the set $ F_{\text{near}} $ are assessed. An eligible facet in $ F_{\text{near}} $ must form an angle with facet $ f_t $ (which contains the sampling site $ \mathbf{s}_i $) such that the absolute cosine of the angle is less than the threshold $ \alpha $. The choice of triangular facet for the second clipping operation depends on its score, with the facet having the lowest score selected as $ f_u $ to clip $ V(\mathbf{s}_i) $. The score for $ f_u $ is determined by two factors: the absolute cosine $ |\cos A| $ of the angle between the eligible facet and $ f_t $, and the Euclidean distance $ \text{disA} $ between the eligible facet and the centroid of $ f_t $:
\begin{equation}
	\text{scoreA} = |\cos A| + \frac{\text{disA}}{d_{\max,i}}
	\label{eq:score nd clip}
\end{equation}
where $ d_{\max,i} $ represents the maximum distance between the neighboring site $ \mathbf{s}_k $ and the Voronoi site $ \mathbf{s}_i $, serving as the normalization term for distance quantification.

Similarly, if a Centroid Voronoi cell $ V(\mathbf{s}_i) $ requires three clippings, all eligible facets in $ F_{\text{near}} $ are evaluated. A facet is eligible if the absolute cosine values of its angles with both $ f_t $ and $ f_u $ exceed the threshold $ \beta $. The selection of $ f_v $ depends on a score computed using three parameters: $ |\cos B| $ (absolute cosine of the angle with $ f_t $), $ |\cos C| $ (absolute cosine with $ f_u $), and their normalized Euclidean distances $ \text{disB}/d_{\max,i} $, $ \text{disC}/d_{\max,i} $. The scoring function is defined as:
\begin{equation}
	 \begin{split}
	\text{scoreB} = |\cos A| + \frac{\text{disA}}{d_{\max,i}} +  |\cos B| \\ +  \frac{\text{disB}}{d_{\max,i}} + |\cos C| + \frac{\text{disC}}{d_{\max,i}}
	\end{split}
	\label{eq:score th clip}
\end{equation}

Notably, while our method implements cross-cell parallelism (i.e., GPU-accelerated computation for distinct Voronoi cells), all sequential clipping operations within individual cells remain strictly non-parallelized.

\begin{figure*}[b!]  % [!htbp] 建议使用更明确的浮动位置参数
	\centering
	\includegraphics[width=0.99\textwidth]{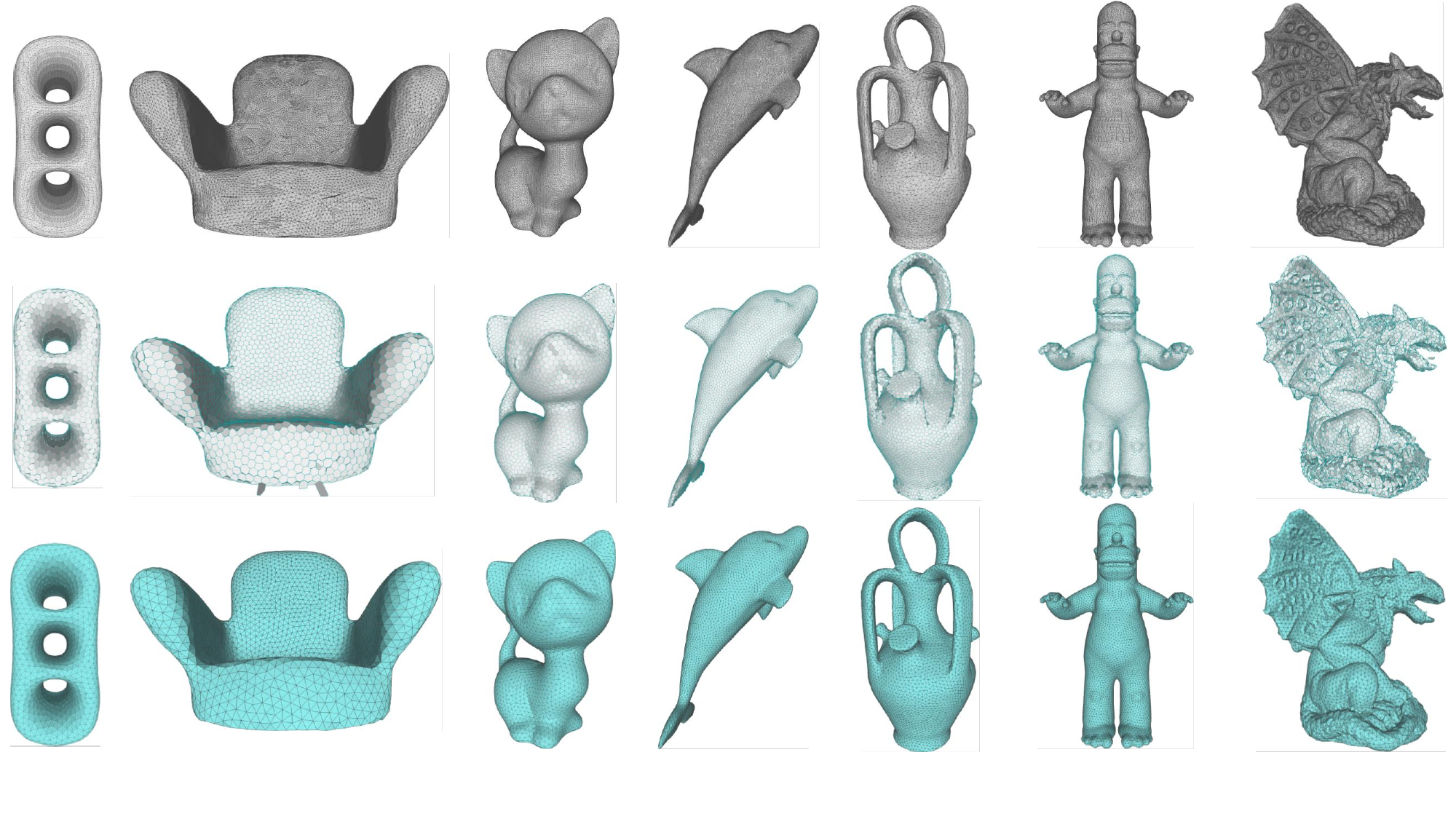} % 建议保留2%边距
	\caption{Validation on seven organic models spanning simple to complex geometries (left to right: Holes, Armchair, Kitten, Dolphin, Botijo, Homer, Gargoyle). Each column shows: (Top) input mesh; (Middle) CVT-optimized intermediate with multi-step clipping; (Bottom) final output mesh. Our method achieves significant triangle count reduction while preserving geometric features and maintaining element quality, even for high-complexity cases (quantitative metrics in \autoref{tab:show_data}).} % 分号分隔层级
	\label{fig:show_fig}  % 正确位置
	\vspace{-2pt}  % 更温和的间距调整
\end{figure*}

\begin{table*}[b!]
	\centering
	\setlength{\tabcolsep}{4pt}
	\captionsetup{labelfont=bf, textfont=it}
	\caption{The quality and running time of various methods on different models correspond to \autoref{fig:show_fig}.}
	\footnotesize
	\label{tab:show_data}
	\begin{tabular}{@{}l l *{12}{c}@{}} % 共 10 列
		\toprule
		\textbf{Model} &  \textbf{In/Out} & \textbf{n} &
		$\bm{Q_{\min}\uparrow}$ & $\bm{Q_{\text{avg}}\uparrow}$ & 
		$\bm{\Theta_{\min}\uparrow}$ & $\bm{\Theta_{\max}\downarrow}$ & 
		$\bm{\Theta_{<30^\circ}\downarrow}$ & $\bm{\Theta_{>90^\circ}\downarrow}$ & 
		$\bm{d_{H} (\times10^{-2})\downarrow}$ & $\bm{RMS(\times10^{-2})\downarrow}$ & $\bm{T(s)\downarrow}$ & 
		$\bm{Q_{\text{up}}~(\%)} \uparrow$ & $\bm{Q_{\text{up}}/T(s)\uparrow}$\\
		\midrule
		% Holes 模型数据
		Holes 
		& Input  & 24.503k   & 0.074 & 0.730 & 3.437 & 167.929 & 0.199 & 0.296 & -- & -- & -- & -- & -- \\
		& Output & 3k & 0.728 & 0.924 & 38.869 & 88.738 & 0.000 & 0.000 & 0.296 & 0.058 & 6.500 & 26.575 & 4.088\\
		\midrule
		Armchair 
		& Input & 39.524k  & 0.024 & 0.697 & 1.257 & 176.288 & 0.259 & 0.391 & -- & -- & -- & -- & -- \\
		& Output & 4k & 0.718 & 0.922 & 38.128 & 89.537 & 0.000 & 0.000 & 0.480 & 0.057 & 6.772 & 32.281 & 4.767 \\
		\midrule
		Kitten 
		& Input  & 55.099k & 0.023 & 0.741 & 1.206 & 176.706 & 0.106 & 0.277 & -- & -- & -- & -- & -- \\
		& Output & 5k & 0.674 & 0.921 & 39.255 & 94.888 & 0.000 & 0.000 & 0.255 & 0.036 & 9.810 & 24.291 & 2.476 \\
		\midrule		
		Dolphin 
		& Input  & 49.181k & 0.245 & 0.781 & 13.949 & 147.587 & 0.064 & 0.255 & -- & -- & -- & -- \\
		& Output & 6k & 0.676 & 0.924 & 37.923 & 93.667 & 0.000 & 0.000 & 0.260 & 0.024 & 7.239 & 18.310 & 2.529 \\
		\midrule
		Botijo 
		& Input  & 10.858k & 0.026 & 0.669 & 1.574 & 176.539 & 0.335 & 0.399 & -- & -- & -- & -- & -- \\
		& Output & 8k & 0.338 & 0.901 & 15.832 & 129.728 & 0.001 & 0.005 & 0.452 & 0.055 & 10.060 & 34.679 & 3.447 \\
		\midrule
		Homer
		& Input  & 31.743k & 0.009 & 0.712 & 0.510 & 178.835 & 0.267 & 0.309 & -- & -- & -- & -- & -- \\
		& Output & 9k & 0.120 & 0.903 & 4.145 & 137.135 & 0.002 & 0.005 & 0.233 & 0.032 & 16.179 & 26.826 &  1.658 \\
		\midrule
		Gargolye 
		& Input  & 100.001k& 0.081 & 0.650 & 2.962 & 169.216 & 0.373 & 0.471 & -- & -- & -- & -- & -- \\
		& Output & 9k & 0.251 & 0.881 & 9.200 & 125.742 & 0.008 & 0.023 & 0.586 & 0.121 & 19.834 & 35.538 & 1.792 \\
		\bottomrule
		\vspace{-3pt}  % 更温和的间距调整
	\end{tabular}
\end{table*}

\subsection{Calculating the centroid}
\label{subsect:method_centroid}
Within our iterative optimization framework, centroids of all clipped cells are recalculated at each iteration to serve as initial sites for subsequent CVT computations.

 As our method specifically targets surface remeshing, these centroids are computed not through the complete geometry of the original Voronoi cells, but via an area-weighted averaging of their clipped polygonal facets. Specifically, we first calculate the geometric centroid $\mathbf{b}_{ij}$ and surface area $A_{ij}$ for each clipped facet. The updated sample point position is then derived by aggregating these values with area-based weighting:
\begin{equation}
	\mathbf{b}_{\text{i}} = \frac{\sum_{i=1}^m A_{ij} \mathbf{b}_{ij}}{\sum_{j=1}^m A_{ij}}
	\label{eq:cal_cent}
\end{equation}
where $m$ represents the total number of clipped facets in the current Voronoi cell. This strategy prioritizes surface-localized geometric features during centroid updates, ensuring alignment with our objective of high-fidelity surface remeshing. As shown in \autoref{fig:calculating_centroid}(\subref{subfig:calculating_centroid_a}).

If the centroid $ \mathbf{b}_{\text{i}} $ does not lie on the original model's surface, it must be projected back to the surface to preserve topological consistency. To achieve this, we adopt a method inspired by RTF \cite{Yao2023_RTF}:  First, the $k$ nearest neighbors of the centroid among the vertices on the original model's surface are identified, denoted as $vn_i^{k}$. Then, the triangles containing these vertices are determined, and the projection distances from the centroid to these triangles are computed. The closest triangle is selected, and the centroid is projected onto it, ensuring alignment with the original surface. This process preserves topological fidelity between the output and original models. Refer to \autoref{fig:calculating_centroid}(\subref{subfig:calculating_centroid_b}).

\subsection{Extracting meshes}
\label{subsect:method_extract}
The RVD(Restricted Voronoi Diagram)-based method in \cite{BOLTCHEVA2017123}, accessible via the Geogram library \cite{Levy2015_Geogram}, is employed for mesh extraction. The optimized sampling points $\mathbf{S}$ serve as direct input to the RVD method, facilitating triangular mesh generation.

\begin{figure*}[t!]
	\centering
	\centering
	% 子图1：使用 subfigure 环境
	\begin{subfigure}{0.245\textwidth}
		\includegraphics[width=\linewidth]{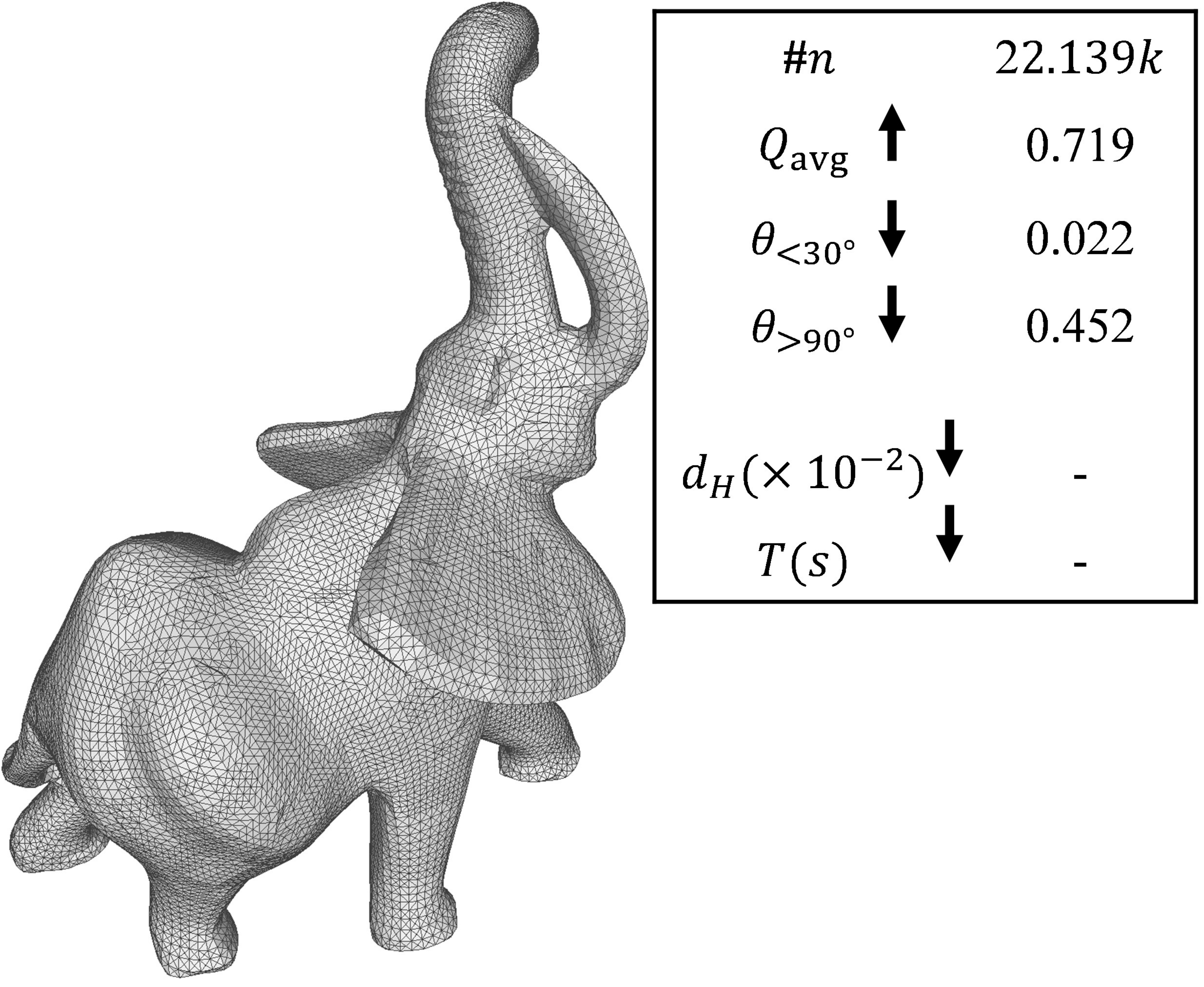}
		\caption{} % 子图标题（留空则不显示编号）
		\label{fig:voronoi_diagram}
	\end{subfigure}
	\hfill
	% 子图2
	\begin{subfigure}{0.245\textwidth}
		\includegraphics[width=\linewidth]{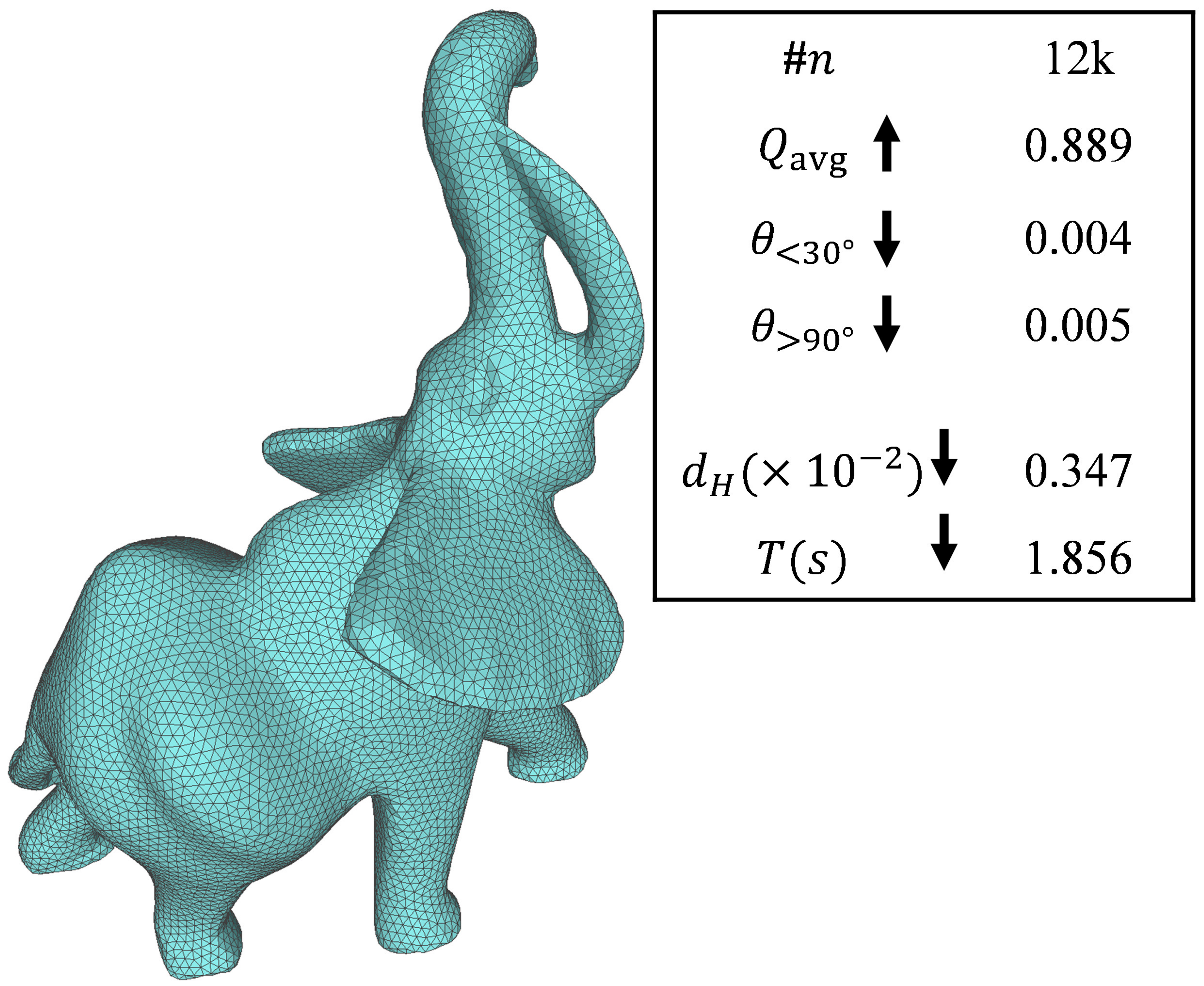}
		\caption{} % 子图标题
		\label{subfig:CVT}
	\end{subfigure}
	% 子图2
	\begin{subfigure}{0.245\textwidth}
		\includegraphics[width=\linewidth]{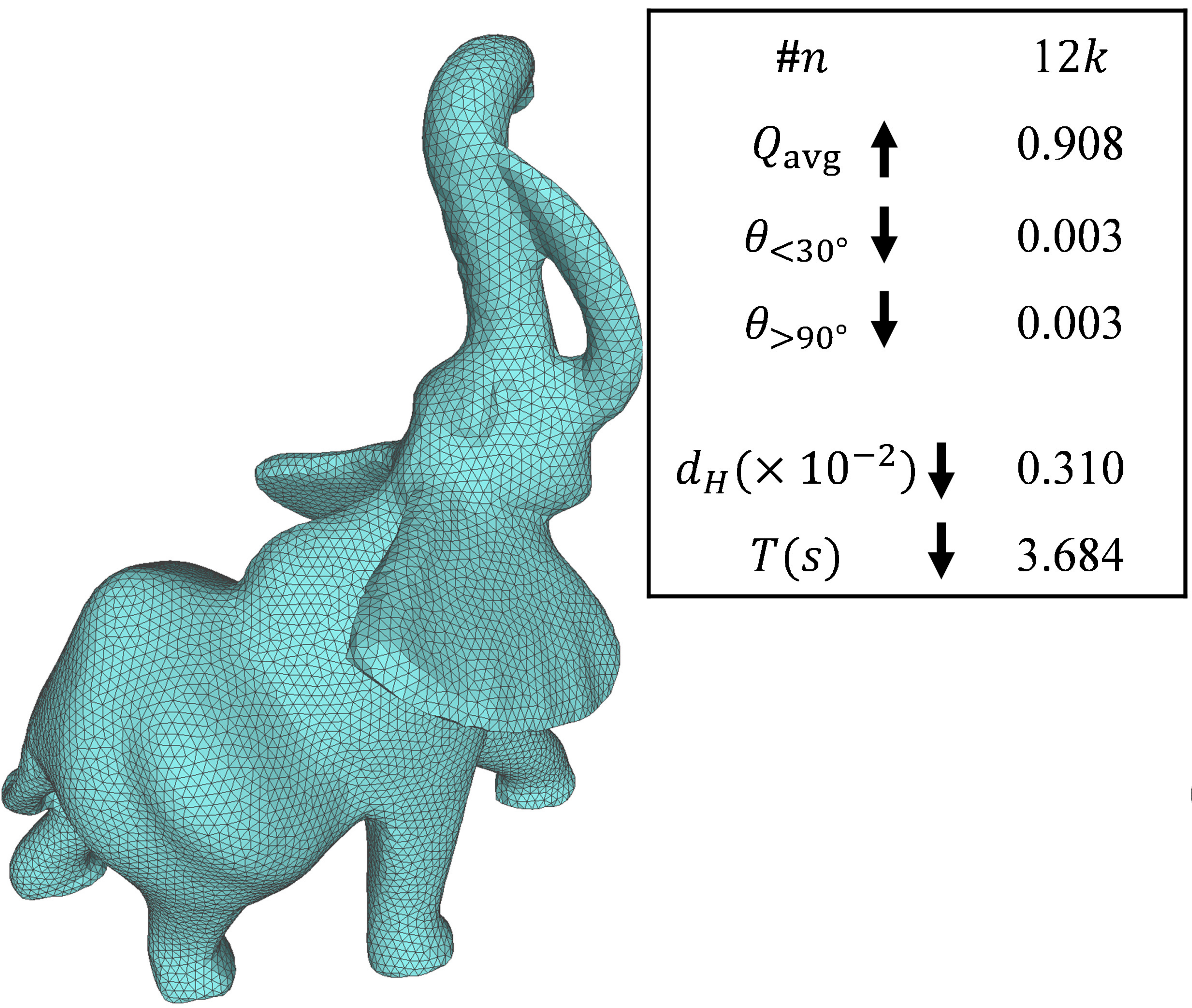}
		\caption{} % 子图标题
		\label{subfig:CVT}
	\end{subfigure}
	% 子图2
	\begin{subfigure}{0.245\textwidth}
		\includegraphics[width=\linewidth]{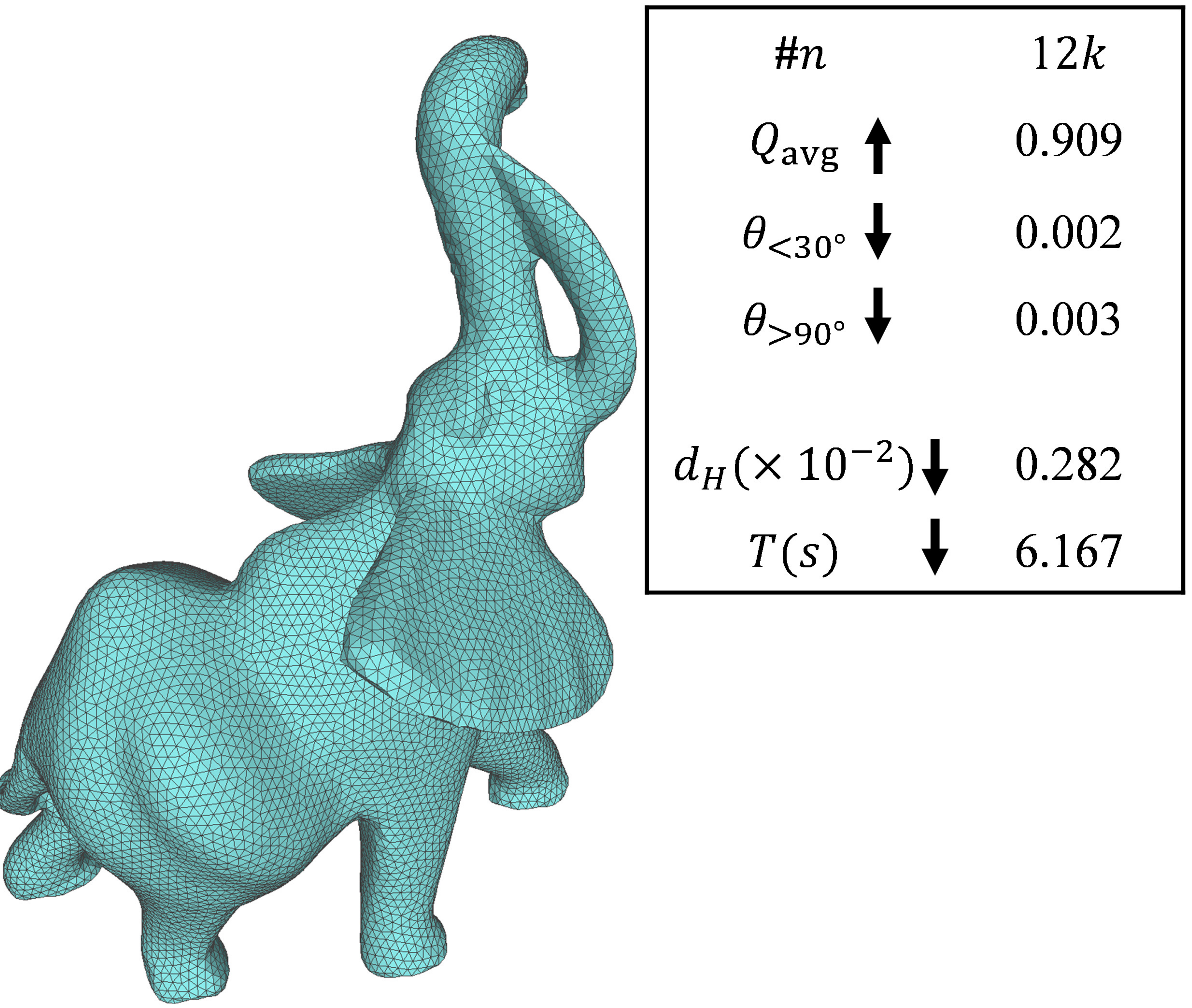}
		\caption{} % 子图标题
		\label{subfig:CVT}
	\end{subfigure}
	\caption{(a) Input model and corresponding triangular mesh quality.
		(b) Remeshed output with maximum clipping times = 1 and its mesh quality.
		(c) Remeshed output with maximum clipping times = 2 and its mesh quality.
		(d) Remeshed output with maximum clipping times = 3 and its mesh quality.
		The results show that when increasing maximum clipping times from 1 to 2, the computational time increases slightly while mesh quality improves significantly. When further increasing times to 3, the computational time grows substantially whereas quality improvement becomes marginal. Our algorithm therefore chooses 3 times as the optimal balance between time cost and quality enhancement.}
	\label{fig:analysis}
	\vspace{-6pt}
\end{figure*}

\begin{figure}[b!]
	\centering
	\centering
	% 子图1：使用 subfigure 环境
	\begin{subfigure}{0.495\linewidth}
		\includegraphics[width=\linewidth]{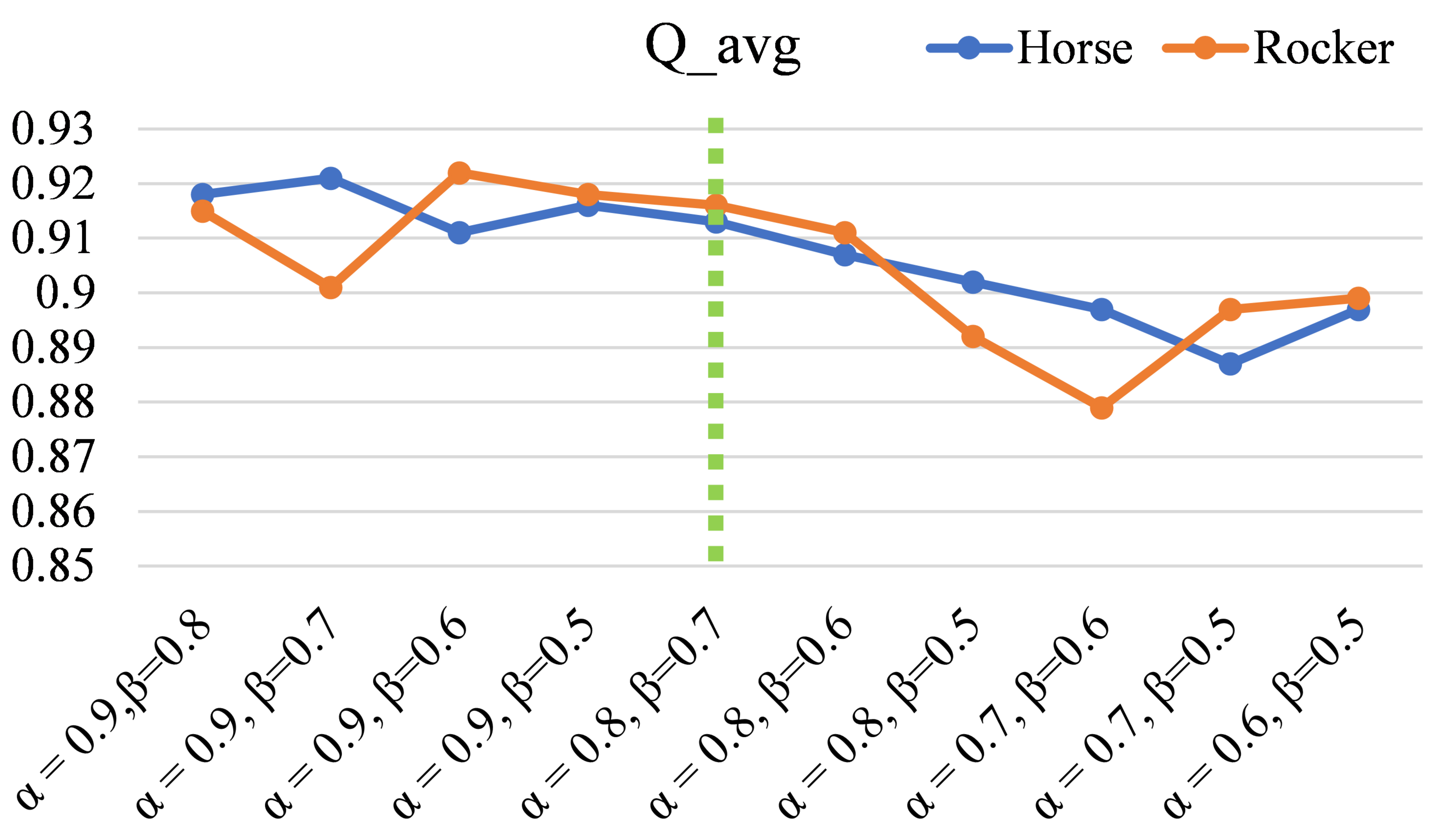}
		\caption{} % 子图标题（留空则不显示编号）
	\end{subfigure}
	\hfill
	% 子图2
	\begin{subfigure}{0.495\linewidth}
		\includegraphics[width=\linewidth]{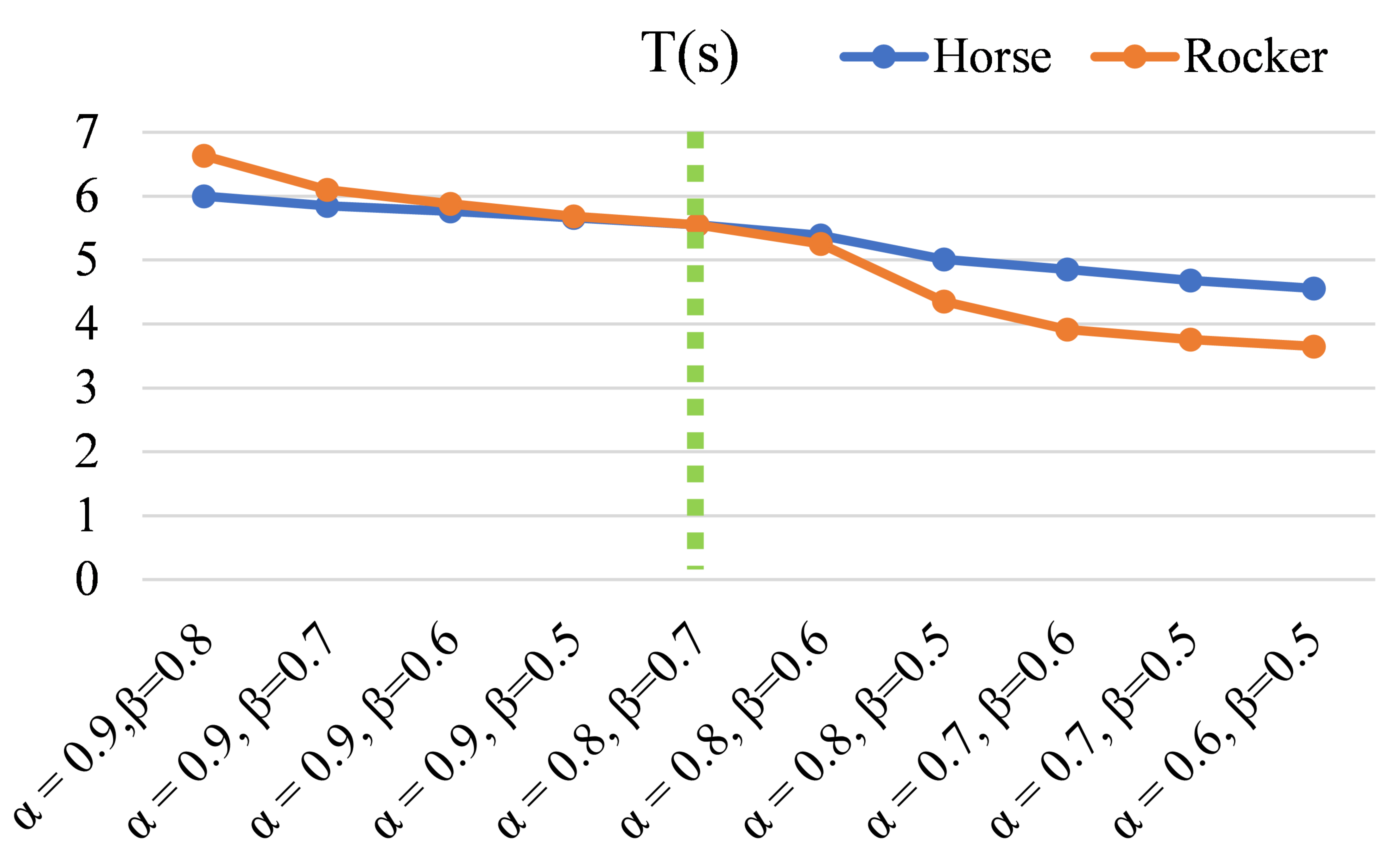}
		\caption{} % 子图标题
	\end{subfigure}
	% 子图2
	\begin{subfigure}{0.495\linewidth}
		\includegraphics[width=\linewidth]{figs/horse.pdf}
		\caption{} % 子图标题
	\end{subfigure}
	% 子图2
	\begin{subfigure}{0.495\linewidth}
		\includegraphics[width=\linewidth]{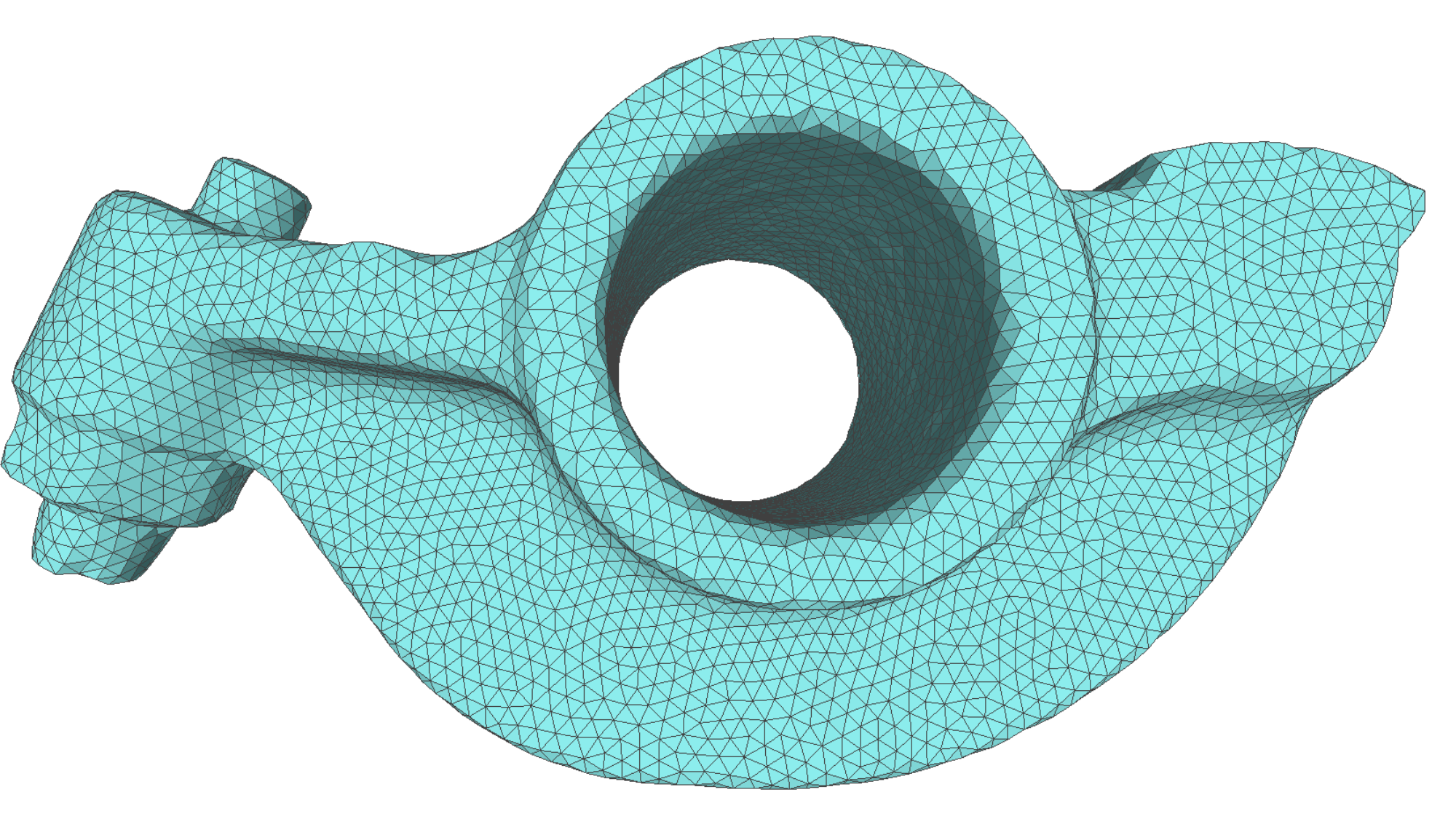}
		\caption{} % 子图标题
	\end{subfigure}
	\caption{(a) $Q_{\text{avg}}$ under different $\alpha$-$\beta$ combinations. 
		(b) $T(s)$ under different $\alpha$-$\beta$ combinations. 
		Experimental results of parameter combinations across different models: Horse (organic model) and Rocker (CAD model). 
		(c) Output model for Horse with $\alpha = 0.8$, $\beta = 0.7$. 
		(d) Output model for Rocker with $\alpha = 0.8$, $\beta = 0.7$. 
		It can be observed that when $\alpha = 0.8$ and $\beta = 0.7$, the output models achieve relatively good quality with moderate time consumption.}
	\label{fig:tunning}
	\vspace{-3pt}
\end{figure}

\section{Experiments}
\label{sect:experiments}
In this section, the effectiveness of the proposed method is validated through experiments using representative models. We conducted comparative experiments with several CVT-based methods, including RVD(Restricted Voronoi Diagram) \cite{Yan2009_RVD}, PowerRTF \cite{10.1145/2999532}, RVC \cite{Chen2018_RVC}, and MPS \cite{Guo2015_MPS}. All experiments were performed on a Windows 11 computer equipped with a 2.1 GHz Intel Core i7-13700 CPU, 32 GB of RAM, and an NVIDIA GeForce RTX 4070 Ti with 12 GB of memory, using CUDA version 12.1. Since the calculations for each cell are independent, the method can be efficiently implemented on GPU.

We evaluate the remeshed models using geometric fidelity and computational efficiency metrics. The angular distribution analysis measures the minimum angle ($\Theta_{\min}$), maximum angle ($\Theta_{\max}$), percentages of angles smaller than $30^\circ$ ($\Theta_{<30^\circ}$), and exceeding $90^\circ$ ($\Theta_{>90^\circ}$). Following \cite{Khan2020_Survey}, three geometric attributes are computed per triangle: area ($A$), semiperimeter ($S$), and longest edge length ($E$). The quality of each triangle is quantified as:
\begin{equation}
	Q = \frac{6}{\sqrt{3}} \cdot \frac{A}{S \cdot E},
\end{equation}
where both $Q_{\min}$ (minimum quality) and $Q_{\text{avg}}$ (average quality) are reported. Global geometric accuracy is evaluated using:
\begin{itemize}
	\item Hausdorff distance $d_H~(\times10^{-2})$ for maximum surface deviation,
	\item Root mean square (RMS) distance $(\times10^{-2})$ for average discrepancy.
\end{itemize}

To quantify efficiency, we propose two metrics:
\begin{itemize}
	\item \textbf{Quality improvement rate}:
	\begin{equation}
		Q_{\text{up}}~(\%) = \left( \frac{Q_{\text{avg}}^{\text{(output)}} - Q_{\text{avg}}^{\text{(input)}}}{Q_{\text{avg}}^{\text{(input)}}} \right) \times 100
	\end{equation}
	
	\item \textbf{Quality improvement per unit time}:
	\begin{equation}
		Q_{\text{up}}/T~(\%\cdot s^{-1}) = \frac{Q_{\text{up}}}{T}
	\end{equation}
\end{itemize}
where $T$ denotes the computation time in seconds. These metrics collectively characterize the method's effectiveness and time-efficiency.

%%%%%%%%%%%%%%%%%%%%%%%%%%%%%%%%%%%

\subsection{Surface Remeshing Results}
The effectiveness and versatility of the proposed surface remeshing method are validated through experimental evaluation. Seven distinct models with varying geometric complexities were selected as input and processed. The visual results of remeshed models are demonstrated in \autoref{fig:show_fig}, with corresponding quantitative metrics provided in \autoref{tab:show_data}. These results confirm the method's consistent performance across different model types.

From the angular distribution perspective, the four simpler models (Holes, Armchair, Kitten, Dolphin) achieve post-remeshing percentages of angles smaller than $30^\circ$ ($\Theta_{<30^\circ}$) and exceeding $90^\circ$ ($\Theta_{>90^\circ}$) approaching zero. For the three complex models, $\Theta_{<30^\circ}$ remains below 0.1.

Regarding triangle quality, all models except the most challenging Gargoyle case attain $Q_{\text{avg}} > 0.9$ after remeshing. Although Gargoyle's $Q_{\text{avg}}$ does not exceed 0.9, it still achieves a quality improvement of 35.54\%.

Geometric fidelity evaluation shows:
\begin{itemize}
	\item Hausdorff distance $d_H < 0.6 (\times10^{-2})$ for all models
	\item RMS distance $<0.06 (\times10^{-2})$ (except Gargoyle at $0.121(\times10^{-2})$)
\end{itemize}

\begin{figure*}[b!]
	\centering
	\includegraphics[width=\linewidth]{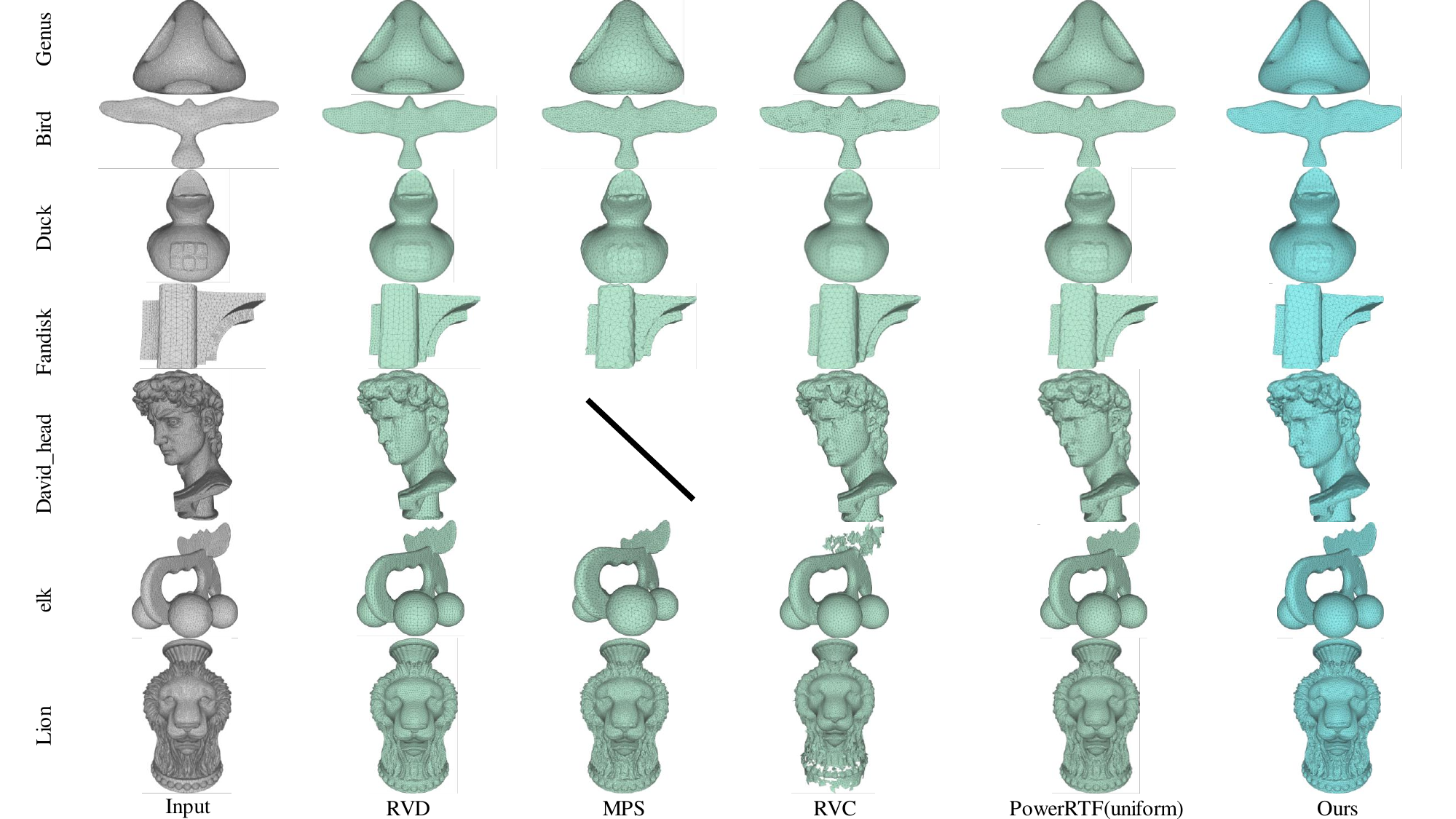}% 图片1
	\caption{Remeshing performance across models with varying geometric complexity (Slash indicates the current method fails to generate remesh results for the corresponding model, i.e., the MPS method cannot produce remesh results for the David\_head model.)}
	\label{fig:compare}
	\vspace{-5pt}
\end{figure*}

\begin{table*}[b!]
	\centering
	\setlength{\tabcolsep}{3pt}
	\captionsetup{labelfont=bf, textfont=it}
	\caption{The quality and running time of various methods on different models correspond to \autoref{fig:compare}. Values in \textcolor{blue}{blue} indicate the best results; values in \textcolor{cyan}{cyan} indicate the second best results.}
	\label{tab:comparison}
	\footnotesize
	\begin{tabular}{@{}l l*{12}{c}@{}} %
		\toprule
		\textbf{Model} & \textbf{Method} & n & $\bm{Q_{\min}{\uparrow}}$ & $\bm{Q_{\text{avg}}{\uparrow}}$ & $\bm{\Theta_{\min}{\uparrow}}$ & $\bm{\Theta_{\max}{\downarrow}}$ & $\bm{\Theta_{<30^\circ}{\downarrow}}$ & $\bm{\Theta_{>90^\circ}{\downarrow}}$ & $\bm{d_{H}(\times10^{-2}){\downarrow}}$ & $\bm{RMS(\times10^{-2})\downarrow}$ & $\bm{T(s)\downarrow}$ & $\bm{Q_{\text{up}}~(\%)} \uparrow$ & $\bm{Q_{\text{up}}/T(\%\cdot s^{-1})\uparrow}$\\ 
		\midrule
		% Genus 模型
		\multirow{5}{*}{Genus}\\
		& Input & 31.421k & 0.053 & 0.771 & 3.429  & 127.991 & 0.078 & 0.289 & -- & --  & -- & --  & --\\
		
		& RVD   & 2.977k & 0.550 & 0.851 & 30.184 & 109.487 & \textbf{\textcolor{blue}{0.000}} & 0.065 & 1.251 & 0.180 & 73.484 & 10.376 & 0.141\\
		
		& MPS   & 3.000k & 0.472 & 0.800 & 27.332 & 119.027 & 0.006 & 0.158 & 1.008 & 0.077 & \textbf{\textcolor{blue}{0.188}} & 3.761 & \textbf{\textcolor{blue}{20.007}}\\
		
		& RVC    & 2.996k & \textbf{\textcolor{blue}{0.632}} & \textbf{\textcolor{cyan}{0.913}} & \textbf{\textcolor{blue}{34.654}} & \textbf{\textcolor{blue}{99.823}} & \textbf{\textcolor{blue}{0.000}} & 0.063 & 1.459 & 0.113  & 3.69 & \textbf{\textcolor{cyan}{18.418}} & 4.991\\
		
		& PowerRTF & 3.000k & 0.485 & 0.867 & 24.322 & 112.105  & \textbf{\textcolor{cyan}{0.002}} & \textbf{\textcolor{cyan}{0.018}} & \textbf{\textcolor{blue}{0.604}} & \textbf{\textcolor{blue}{0.060}} & \textbf{\textcolor{cyan}{1.72}} & 12.451 & \textbf{\textcolor{cyan}{7.239}}\\
		
		& Ours   & 3.000k & \textbf{\textcolor{cyan}{0.586}} & \textbf{\textcolor{blue}{0.916}} & \textbf{\textcolor{cyan}{30.460}} & \textbf{\textcolor{cyan}{101.487}} & \textbf{\textcolor{blue}{0.000}} & \textbf{\textcolor{blue}{0.001}} & \textbf{\textcolor{cyan}{0.783}} & \textbf{\textcolor{cyan}{0.068}} & 6.500 & \textbf{\textcolor{blue}{18.807}} & 2.893\\ 
		\midrule
		% Bird 模型
		\multirow{5}{*}{Bird} 
		& Input  & 2.431k & 0.414 & 0.807 & 24.864 & 126.238 & 0.010 & 0.157 & -- & --  & -- & --  & --\\
		
		& RVD   & 3.043k & \textbf{\textcolor{cyan}{0.546}} & 0.853 & \textbf{\textcolor{cyan}{28.964}} & \textbf{\textcolor{cyan}{109.877}} & \textbf{\textcolor{blue}{0.000}} & 0.048 & 0.613 & 0.120 & 51.084 & 5.700 & 0.112 \\
		
		& MPS    & 3.003k & 0.447 & 0.792 & 25.798 & 122.041 & 0.012 & 0.180 & 0.352 & 0.093 & \textbf{\textcolor{cyan}{0.125}} & nan & nan\\
		
		& RVC    &2.423k & 0.373 & \textbf{\textcolor{cyan}{0.887}} & 15.487  & 113.139 & 0.004 & 0.027 & 1.488 & 0.368 & 2.551 & \textbf{\textcolor{cyan}{9.913}} & 3.886\\
		
		& PowerRTF & 3.000k & 0.499 & 0.865 & 23.968 & 110.545 & \textbf{\textcolor{cyan}{0.003}} & \textbf{\textcolor{cyan}{0.023}} & \textbf{\textcolor{cyan}{0.336}} & \textbf{\textcolor{cyan}{0.081}} & 1.646 & 7.187 & \textbf{\textcolor{cyan}{4.369}} \\
		
		& Ours   & 3.000k & \textbf{\textcolor{blue}{0.582}} & \textbf{\textcolor{blue}{0.915}} & \textbf{\textcolor{blue}{30.635}} & \textbf{\textcolor{blue}{102.412}} & \textbf{\textcolor{blue}{0.000}} & \textbf{\textcolor{blue}{0.002}} & \textbf{\textcolor{blue}{0.331}} & \textbf{\textcolor{blue}{0.075}} & \textbf{\textcolor{cyan}{1.169}} & \textbf{\textcolor{blue}{13.383}} & \textbf{\textcolor{blue}{11.448}}\\ 
		\midrule
		
		% Duck 模型
		\multirow{5}{*}{Duck} 
		& Input  & 24.988k & 0.030 & 0.756 & 1.583  & 175.713 & 0.140 & 0.259 & -- & --  & -- & --  & --\\
		
		& RVD   & 2.934k & 0.530 & 0.849 & 28.884 & 111.857 & \textbf{\textcolor{blue}{0.000}} & 0.063 & 0.693 & 0.095 & 69.274 & 12.302 & 0.178 \\
		
		& MPS    & 3.001k & 0.473 & 0.800 & 27.705 & 118.789  & 0.004 & 0.165 & 0.795 & 0.085 & \textbf{\textcolor{blue}{0.156}} & 5.820 & \textbf{\textcolor{blue}{37.308}}\\
		
		& RVC    & 2.999k & \textbf{\textcolor{cyan}{0.583}} & \textbf{\textcolor{cyan}{0.908}} & \textbf{\textcolor{blue}{30.067}} & \textbf{\textcolor{cyan}{105.584}}  & \textbf{\textcolor{cyan}{0.001}} & \textbf{\textcolor{cyan}{0.009}} & 0.755 & 0.102 & \textbf{\textcolor{cyan}{2.084}} & \textbf{\textcolor{cyan}{20.106}} & \textbf{\textcolor{cyan}{9.648}}\\
		
		& PowerRTF & 3.000k & 0.534 & 0.869 & 26.241 & 105.781 & 0.002 & 0.016 & \textbf{\textcolor{blue}{0.432}} & \textbf{\textcolor{blue}{0.066}} & 2.463 & 13.003 & 5.280\\
		
		& Ours   & 3.000k & \textbf{\textcolor{blue}{0.633}} & \textbf{\textcolor{blue}{0.917}} & \textbf{\textcolor{cyan}{29.717}} & \textbf{\textcolor{blue}{96.923}} & \textbf{\textcolor{blue}{0.000}} & \textbf{\textcolor{blue}{0.001}} & \textbf{\textcolor{cyan}{0.536}} & \textbf{\textcolor{cyan}{0.073}} & 3.436 & \textbf{\textcolor{blue}{21.296}} & 6.198 \\ 
		\midrule
		
		% Fandisk 模型
		\multirow{5}{*}{Fandisk} 
		& Input  & 7.223k & 0.179 & 0.625 & 0.759 & 176.536 & 0.481 & 0.310 & -- & --  & -- & -- & --\\
		
		& RVD   & 3.156k & 0.522 & 0.829 & \textbf{\textcolor{cyan}{28.590}} & 112.615 & \textbf{\textcolor{blue}{0.000}} & 0.066 & 0.929 & 0.169 & 45.385 & 32.640 & 0.719 \\
		
		& MPS    & 3.006k & 0.410 & 0.798 & 25.341 & 126.711 & 0.009 & 0.166 & 1.219 & 0.165 & \textbf{\textcolor{blue}{0.125}} & 27.68 & \textbf{\textcolor{blue}{221.44}}\\
		
		& RVC    & 2.993k & 0.394 & \textbf{\textcolor{cyan}{0.896}} & 22.059  & 128.175  & 0.002 & \textbf{\textcolor{cyan}{0.016}} & 1.999 & 0.320 & 3.097 & \textbf{\textcolor{cyan}{43.360}} & 14.001\\ 
		
		& PowerRTF & 3.000k & \textbf{\textcolor{cyan}{0.538}} & 0.864 & 25.226 & \textbf{\textcolor{cyan}{110.61}} & \textbf{\textcolor{cyan}{0.001}} & 0.022 & \textbf{\textcolor{blue}{0.747}} & \textbf{\textcolor{blue}{0.111}} & \textbf{\textcolor{cyan}{2.315}} & 38.240 & 16.518\\
		
		& Ours   & 3.000k & \textbf{\textcolor{blue}{0.663}} & \textbf{\textcolor{blue}{0.912}} & \textbf{\textcolor{blue}{35.742}} & \textbf{\textcolor{blue}{95.913}} & \textbf{\textcolor{blue}{0.000}} & \textbf{\textcolor{blue}{0.002}} & \textbf{\textcolor{cyan}{0.824}} & \textbf{\textcolor{cyan}{0.143}} & 2.672 & \textbf{\textcolor{blue}{45.92}} & \textbf{\textcolor{cyan}{17.186}}\\ 
		\midrule
		
		% David head 模型
		\multirow{5}{*}{David head} 
		& Input  & 108.333k & 0.046 & 0.654 & 2.455  & 173.76 & 0.361 & 0.461 & -- & -- & --  & -- & --\\
		
		& RVD & 6.114k & \textbf{\textcolor{blue}{0.518}} & 0.849 & \textbf{\textcolor{blue}{25.369}} &\textbf{\textcolor{blue}{112.322}} & \textbf{\textcolor{blue}{0.000}} & 0.058 & \textbf{\textcolor{blue}{0.602}} & 0.158 & 392.818 & \textbf{\textcolor{cyan}{29.817}} & 0.076\\
		
		& MPS    & -- & -- & --  & -- & -- & -- & -- & -- & --  & -- & -- & --\\
		
		& RVC    &5.917k & \textbf{\textcolor{cyan}{0.336}} & \textbf{\textcolor{cyan}{0.890}} & \textbf{\textcolor{cyan}{13.456}}  & 116.225  & \textbf{\textcolor{cyan}{0.002}} & \textbf{\textcolor{cyan}{0.024}} & 2.114 & 0.183 & \textbf{\textcolor{cyan}{7.647}} & 36.086 & \textbf{\textcolor{cyan}{4.719}} \\ 
		
		& PowerRTF & 6.000k & 0.094 & 0.867 & 3.250 & 130.014 & 0.007 & 0.025 & 0.772 & \textbf{\textcolor{cyan}{0.136}} & \textbf{\textcolor{blue}{5.316}} & \textbf{\textcolor{blue}{32.569}} & \textbf{\textcolor{blue}{6.127}}\\
		
		& Ours   & 6.000k & 0.182 & \textbf{\textcolor{blue}{0.896}} & 6.459 & \textbf{\textcolor{cyan}{115.713}} & 0.003 & \textbf{\textcolor{blue}{0.014}} & \textbf{\textcolor{cyan}{0.667}} & \textbf{\textcolor{blue}{0.124}} & 10.512 & 37.003 & 3.520\\ 
		
		% Elk 模型
		\midrule
		\multirow{5}{*}{Elk} 
		& Input  & 10.716 & 0.019 & 0.821 & 0.863 & 176.633 & 0.051 & 0.101 & -- & -- & --  & -- & --\\
		
		& RVD &7.146k & \textbf{\textcolor{blue}{0.518}} & 0.849 & \textbf{\textcolor{cyan}{25.369}} & \textbf{\textcolor{blue}{112.322}} & \textbf{\textcolor{blue}{0.000}} & 0.058 & \textbf{\textcolor{cyan}{0.567}} & 0.093 & 226.232 & \textbf{\textcolor{blue}{11.858}} & 0.052\\
		
		& MPS   & 7.000k & 0.416 & 0.795 & 25.363 & 125.963 & 0.008 & 0.178 & 0.637 & 0.078 & \textbf{\textcolor{blue}{0.297}} & nan & nan\\
		
		& RVC   & 6.860k & 0.022 & \textbf{\textcolor{cyan}{0.895}} & 0.746  & 174.673  & 0.013 & 0.027 & 2.998 & 0.349 & 24.922 & 9.013 & 0.362\\ 
		
		& PowerRTF & 7.000k & 0.471 & 0.864 & 22.656 & 116.628 & \textbf{\textcolor{cyan}{0.004}} & \textbf{\textcolor{cyan}{0.023}} & \textbf{\textcolor{blue}{0.516}} & \textbf{\textcolor{cyan}{0.071}} & 2.97 & 5.238 & \textbf{\textcolor{cyan}{1.764}}\\
		
		& Ours   & 7.000k & \textbf{\textcolor{cyan}{0.495}} & \textbf{\textcolor{blue}{0.916}} & \textbf{\textcolor{blue}{27.483}} & \textbf{\textcolor{cyan}{115.19}} & \textbf{\textcolor{blue}{0.000}} & \textbf{\textcolor{blue}{0.002}} & 0.594 & \textbf{\textcolor{blue}{0.066}} & \textbf{\textcolor{cyan}{3.47}} & \textbf{\textcolor{cyan}{11.571}} & \textbf{\textcolor{blue}{3.335}} \\ 
		
		% Lion 模型
		\midrule
		\multirow{5}{*}{Lion} 
		& Input  & 40k & 0.307 & 0.496 & 1.283 & 175.596 & 0.658 & 0.682  & -- & -- & --  & -- & --\\
		
		& RVD & 20.266k & \textbf{\textcolor{cyan}{0.284}} & 0.851 & \textbf{\textcolor{cyan}{14.205}} & \textbf{\textcolor{cyan}{141.66}} & \textbf{\textcolor{blue}{0.006}} & 0.064 & \textbf{\textcolor{blue}{0.559}} & \textbf{\textcolor{cyan}{0.114}} & 1348.96 & 71.573 & 0.053\\
		
		& MPS    & 20.028k & \textbf{\textcolor{blue}{0.368}} & 0.790 & \textbf{\textcolor{blue}{21.824}} & \textbf{\textcolor{blue}{131.836}} & 0.020 & 0.181 & 0.632 & 0.118 & \textbf{\textcolor{blue}{1.093}} & 59.274 & \textbf{\textcolor{blue}{54.231}}\\
		
		& RVC    & 17.269k & 0.103 & 0.858 & 5.295  & 166.344  & 0.015 & 0.056 & 5.114 & 0.235 & \textbf{\textcolor{cyan}{6.265}} & 72.984 & \textbf{\textcolor{cyan}{11.649}}\\
		
		& PowerRTF & 20.000k & 0.075 & \textbf{\textcolor{cyan}{0.881}} & 2.557 & 159.706 & 0.014 & \textbf{\textcolor{cyan}{0.030}} & 0.885 & 0.128 & 9.197 & \textbf{\textcolor{cyan}{77.621}} & 8.440\\
		
		& Ours   & 20.000k & 0.072 & \textbf{\textcolor{blue}{0.891}} & 2.451 & 149.668 & \textbf{\textcolor{cyan}{0.009}} & \textbf{\textcolor{blue}{0.023}} & \textbf{\textcolor{cyan}{0.620}} & \textbf{\textcolor{blue}{0.094}} & 9.514 & \textbf{\textcolor{blue}{79.637}} & 8.371\\ 
		\bottomrule
		\vspace{-2pt}
	\end{tabular}
\end{table*}

Computational analysis reveals that repeated clipping operations on high-curvature regions (e.g., sharp edges and hair-like structures) increase time consumption for complex models. This leads to reduced quality improvement per unit time ($Q_{\text{up}}/T$) despite high absolute quality gains. Nevertheless, the method maintains effective geometric preservation across all test cases.

This comprehensive validation demonstrates robust capability in handling complex geometries while maintaining stable convergence properties.

\subsection{Parameter Analysis}\label{sect:analysis}
\textbf{\textit{Maximum clipping times and remeshing quality.}}
In our method, the maximum number of clipping times has a decisive influence on the quality of remeshing. We conducted comparative experiments by limiting the maximum clipping times to 1, 2, and 3. The results (shown in \autoref{fig:analysis}) demonstrate that when the number of times is set to 1, the remeshing quality reaches the baseline level; increasing to 2 times yields a small improvement in quality; and allowing 3 times leads to further enhancement of quality metrics.

When all clipping times are set to 1, the computational pattern can be regarded as similar to the fast approximation method PowerRTF \cite{https://doi.org/10.1111/cgf.14897}. When the number of clipping times is set to higher values, the method gradually approaches the precision of the exact intersection method RVD \cite{Yan2009_RVD}. This adjustable parameter characteristic enables our method to cover various application scenarios ranging from efficiency-priority to precision-priority, reflecting the flexibility advantage of the framework.

\textbf{\textit{Curvature - based threshold selection for adaptive clipping.}}
In this approach, the curvature of the position of the Voronoi cells plays a critical role in determining whether multiple clips are necessary. Given the challenges of accurately calculating curvature for complex models, a simplified estimation method is implemented that approximates curvature by evaluating the angles between adjacent faces. The cosine values of these angles are normalized to a range of $[0, 1]$, and thresholds $\alpha$ and $\beta$ are established. When the normalized value falls below these thresholds, it indicates that the angle is suitable for performing either two or three clips.

Through thorough analysis, appropriate threshold values are identified. Smaller thresholds effectively reduce the number of Voronoi cells requiring multiple clips, thereby decreasing computational complexity, though this may compromise the quality of the results. Conversely, larger thresholds increase the number of Voronoi cells being clipped, which improves overall quality but also elevates computational demands. After carefully balancing these considerations through experimentation, the selected thresholds are $\alpha = 0.8$ and $\beta = 0.7$, with the corresponding experimental data presented in \autoref{fig:tunning}.
%%%%%%%%%%%%%%%%%%%%%%%%%%%%%%%%%%%

%%%%%%%%%%%%%%%%%%%%%%%%%%%%%%%%%%%

\subsection{Comparison}\label{sect:comparison}
Several recent CVT-based methods for surface remeshing have been selected for comparison. The results show in \autoref{fig:compare}. Quantitative evaluation metrics are provided in \autoref{tab:comparison}. Our method consistently achieves optimal $Q_{\text{avg}}$ values, indicating significant improvement in overall remeshing quality. For $\Theta_{<30^\circ}$ and $\Theta_{>90^\circ}$, it maintains optimal or suboptimal values, confirming effective control of undersized and oversized angles. The $d_{\text{H}}~(\times10^{-2})$ and $\mathrm{RMS}~(\times10^{-2})$ results surpass most baselines, demonstrating precise geometric preservation.

The results exhibit strong robustness across diverse models, including simple-to-complex structures and smooth-to-sharp features. Comparative analysis between the exact intersection method (Restricted Voronoi Diagram \cite{Yan2009_RVD}) using original facets and the approximate tangent facets method (PowerRTF \cite{10.1145/2999532}) with tangent planes is performed in terms of time $T(s)$ versus $Q_{\text{avg}}$. 

Our adaptive clipping strategy applies multiple clips to Voronoi cells in high-curvature regions (sharp edges, detailed textures) and fewer clips in low-curvature areas. This results in faster processing for simple models (smooth surfaces, low curvature) and longer computation for complex cases (rough surfaces, high curvature variations). Nevertheless, both $Q_{\text{up}}~(\%)$ and $Q_{\text{up}}/T~(\%\cdot s^{-1})$ metrics confirm our method's high stability and consistent efficiency in quality enhancement per unit time.

\section{Conclusion}
\label{sect:conclusion}
This paper proposes a CVT-based remeshing algorithm. It neither precisely computes the intersection between 3D CVT and the original model facets nor simply approximates complex surfaces using planes. Instead, it dynamically balances computational complexity and output mesh quality by adjusting the number of clipping times (i.e., computing intersections between 3D CVT and corresponding original facets) based on the curvature of original facets where CVT sites reside. Multiple experimental results demonstrate the effectiveness of our method.

\textit{Limitations and future work}
Although our method performs well on various models, some limitations remain. First, it employs uniform remeshing, while practical models require denser sampling points in high-curvature areas (e.g., lion’s mane) than flat regions (e.g., lion’s base) to better fit details. A curvature-adaptive remeshing approach might yield improved results. Second, we do not preserve features, making it unsuitable for CAD models requiring strong feature retention. Finally, our results may depend on the original model quality.

In future work, we will incorporate curvature-based sampling for CVT site distribution and introduce feature preservation to better capture model details and sharp facets.

%%Vancouver style references.
\bibliographystyle{cag-num-names}
\bibliography{refs}

\end{document}